\documentclass[prd,showpacs,showkeys,preprintnumbers,floatfix,
nofootinbib,superscriptaddress,twocolumn]{revtex4-1}
\usepackage{float}
\usepackage{nicefrac}
\usepackage{mathtools}
\usepackage{amsfonts} % AMS
\usepackage{amssymb} % AMS
\usepackage{amsmath} % AMS
\usepackage{graphicx} % Include figure files
\usepackage{subfigure} % Include figure files
\usepackage{array} % array
\usepackage{dcolumn} % Align table columns on decimal point
\usepackage{bm} % bold math
\usepackage{latexsym} % latex symbols
\usepackage{longtable} % long tables
\usepackage{hyperref} % hypertext links 
\usepackage{verbatim}
\usepackage{epsfig}
\usepackage{color}
\newcommand{\sh}[0]{\,\mathrm{sh}}
\newcommand{\Kdf}[0]{\mathcal K_{\mathrm{df},3}}

\newcommand{\Mth}[0]{\mathcal M_{ 3}}
\newcommand{\ML}[0]{\mathcal M_{ 2,L}}
\newcommand{\MthL}[0]{\mathcal M_{ 3,L}}

\newcommand{\M}[0]{\mathcal M_2}

\begin{document}
\title{Applying the relativistic quantization condition to a\\ three-particle bound state in a periodic box}
\author{Maxwell T. Hansen}
\email[e-mail: ]{hansen@kph.uni-mainz.de}
\affiliation{
Institut f\"ur Kernphysik and Helmholtz Institute Mainz, Johannes Gutenberg-Universit\"at Mainz,
55099 Mainz, Germany\\
}
\author{Stephen R. Sharpe}
\email[e-mail: ]{srsharpe@uw.edu}
\affiliation{
 Physics Department, University of Washington, 
 Seattle, WA 98195-1560, USA \\
}
\date{\today}
\begin{abstract}
Using our recently developed relativistic three-particle quantization
condition~\cite{SpectoK,KtoM}, we study the finite-volume energy shift
of a spin-zero three-particle bound state. We reproduce the result obtained using
non-relativistic quantum mechanics by Mei{\ss}ner, R\'ios and Rusetsky~\cite{MRR},
and generalize the result to a moving frame.

\end{abstract}
%
%\pacs{11.80.-m, 11.80.Jy, 11.80.La, 12.38.Gc }
%
\keywords{finite volume, lattice QCD}
\maketitle

\section{Introduction}

There has been considerable recent progress using lattice QCD to study
resonances (as reviewed, 
for example, in Refs.~\cite{Lang:2015ljt,Edwards:2016gku,Dudek:2016bxq}).
This is mainly based on a line of theoretical work, begun 
by L\"uscher in Refs.~\cite{Luscher:1986pf,Luscher:1991cf}, 
that relates the spectrum of multiple-particle
states in a finite volume (FV) to infinite-volume scattering amplitudes.
Until recently, this work has been restricted to resonances (or bound states) 
that couple only to two-particle channels.
Since many resonances and bound states couple to channels containing more than
two particles, it is necessary to extend the theoretical formalism to three or more
particles.

Recently, we derived a generalization of L\"uscher's work that applies for three
identical, spinless relativistic particles~\cite{SpectoK,KtoM}. 
Specifically, we obtained a quantization condition that relates three-particle
energies in a cubic box of size $L$ to the two-particle scattering amplitude $\M$
and an infinite-volume three-particle scattering K-matrix, $\Kdf$,
as well as a set of integral equations relating $\Kdf$ and $\M$ to the physical
three-particle scattering amplitude $\Mth$.
Our result assumes that the two-particle K-matrix has no poles on the real energy axis in the kinematic region of interest,
and also assumes a symmetry that decouples even- and odd-particle-number states. 
The first restriction must be imposed because two-particle K-matrix poles give rise to finite-volume effects that we did not include in our derivation. The second restriction reduces the classes of diagrams that contribute and thus simplifies the derivation.

Other than this, the result is completely general.
Both the derivation and the final expressions are, 
however, rather complicated, and it is important to provide cross-checks
of the formalism. We have completed one such check in Ref.~\cite{thresh} by comparing
the FV energies of the state nearest to the three-particle threshold to results 
obtained using non-relativistic quantum mechanics 
(NRQM)~\cite{Huang:1957im,Beane:2007qr,Tan:2007}
and relativistic perturbation theory~\cite{HSpert}.

The purpose of this paper is to provide another, completely independent check 
on the formalism,
by using it to determine the leading volume dependence of the binding
energy of a spin-zero three-particle bound state.
Using NRQM, Mei{\ss}ner, R\'ios and Rusetsky (MRR) have calculated this dependence in Ref.~\cite{MRR}. In that work, the authors restrict attention to a system with two-particle interactions near the unitary limit, 
so that Efimov-like three-particle bound states appear~\cite{Efimov:1970zz}. 
Here we determine the leading energy dependence for the same system,
using our relativistic formalism, and find complete
agreement with the NRQM result.

The derivation of this result in our formalism is quite involved.
In particular, as noted above, the quantization condition depends on the intermediate, regularization-dependent quantity, $\Kdf$, whereas  the final result for the 
 energy shift must depend only on physical quantities.
Seeing how this happens 
gives us insight into the workings of the formalism.

MRR consider the case of a bound state at rest {in the finite volume}.
It has been found for two-particle bound states that the leading volume dependence
can be canceled by combining results for bound states with differing total 
momenta, $\vec P$~\cite{Davoudi:2011md}.
Thus it is interesting to generalize the three-particle analysis also to moving bound states.
It turns out that our derivation of the finite-volume energy shift can readily be generalized 
to $\vec P\ne 0$, {as we describe in Sec.~\ref{sec:moving}}.

The remainder of this article is organized as follows.
In the next section we describe the result of MRR.
Then, in Sec.~\ref{sec:QCexp},
we use our quantization condition to 
derive a general prediction for the leading-order energy shift, $\Delta E(L)$,
in terms of unsymmetrized versions of the residue factors
(which are the on-shell limit of unsymmetrized Bethe-Salpeter amplitudes).
This section is the core of the paper.
Next, in Sec.~\ref{sec:calcGamma},
we relate the residue factors to the components of the Faddeev wavefunction in the
NRQM analysis.
With these results in hand, in Sec.~\ref{sec:calcDE} 
we evaluate our expression for $\Delta E(L)$,
finding the MRR result. 
In Sec.~\ref{sec:twopart},
we briefly compare our analysis with that for a two-particle bound state,
and then, in Sec.~\ref{sec:moving}, we
discuss the generalization to nonzero total momentum for both 
two- and three-particle  bound states.
We conclude in Sec.~\ref{sec:conc}.
Technical details are relegated to three appendices.
In the first we explain why several approximations made in the main text do not
impact the leading-order volume dependence.
In the second we relate the on-shell Bethe-Salpeter amplitudes to the
three-particle Schr\"odinger wavefunction.
In the final appendix we derive an identity for the Schr\"odinger wavefunction of the Efimov state.

\section{Preliminaries and the MRR result}

\label{sec:setup}

Following MRR we consider identical spinless scalar particles (of mass $m$) 
in the unitary limit in which the two-particle s-wave scattering length diverges. 
We adopt the convention $p \cot \delta(p) = - 1/a + \cdots$, and take the scattering length, $a$,
to be negative so that there are no two-particle
bound states near threshold.
It is well known that such a system has a tower of three-particle
bound states, known as Efimov states~\cite{Efimov:1970zz}.
Focusing attention on any one of these states, 
we write the associated pole in the infinite-volume
three-to-three scattering amplitude as
\begin{equation}
\label{eq:pole}
\mathcal M_3(\vec p, \hat a'^*; \vec k, \hat a^*) 
\sim - \frac{\Gamma(\vec p, \hat a'^*) 
\overline \Gamma( \vec k, \hat a^*)}{E^{*2} - E_B^{2} } \,.
\end{equation}
Here $\mathcal M_3$ is a function of center-of-mass (CM) frame energy, $E^*$, 
as well as two copies of on-shell three-particle phase space 
(the parametrization of which will be explained below).
The $\sim$ indicates that the difference of the two sides is finite at the pole. 
We also introduce the binding momentum $\kappa$, defined by
\begin{align}
E_B 
& = 3 m - \frac{\kappa^2}{m}\,.
\end{align}
Following MRR, we assume a shallow bound state, so that $\kappa \ll m$.

The residue of the pole is determined by the matrix elements 
\begin{align}
\label{eq:Gammadef}
(2 \pi)^4 \delta^4(P-P_B) i \Gamma(\vec p, \hat a'^*) 
&\equiv  \langle 3 \phi, \mathrm{out} \vert E_B \rangle \,,
\\
\label{eq:Gammabardef}
(2 \pi)^4 \delta^4(P-P_B) i \overline\Gamma(\vec k, \hat a^*) 
&\equiv  \langle E_B\vert  3 \phi, \mathrm{in} \rangle \,,
\end{align}
between the bound state and the three-particle asymptotic states, 
analytically continued below threshold. Here $P$ is the four-momentum of the three-particle states and $P_B$ the four-momentum of the bound state. 
The states in Eqs.~(\ref{eq:Gammadef}) and (\ref{eq:Gammabardef}) have standard, relativistic normalization, so that $\Gamma$ and $\overline \Gamma$ are
dimensionless. These quantities are functions of on-shell three-particle phase space
evaluated at the fixed subthreshold CM energy $E_B$. 

In the above we have used the
coordinate system for three on-shell particles introduced in
Ref.~\cite{SpectoK}. Specifically, we consider three particles with fixed
total energy and momentum, $E$ and $\vec P$. Although our quantization
condition holds for general total momentum, for most of this work we restrict attention
to $\vec P = 0$, since this is the case studied by MRR.
Thus there is no distinction between CM-frame and moving-frame energies, 
so we use $E$ rather than $E^*$ in the following sections. 
We relax this restriction in Section~\ref{sec:moving},
where we consider nonzero total momentum.

To specify the coordinate system we fix the momentum of one of the three particles
to be $\vec k$ (the ``spectator momentum''), and, since the particle is on-shell, its energy is
\begin{equation}
\omega_k = \sqrt{k^2 + m^2} \,,
\end{equation}
where $k \equiv \vert \vec k \vert$. 
The total energy-momentum of the other two particles is 
then constrained to be $(E - \omega_k, - \vec k)$.
In their CM frame, these two particles thus have total energy
\begin{equation}
E_{2,k}^*  = \sqrt{(E - \omega_k)^2 - k^2} \,, 
\label{eq:E2kdef}
\end{equation}
and individual momenta 
\begin{equation}
q_k^*  = \sqrt{E_{2,k}^{*2}/4-m^2} \,.
\label{eq:qkdef}
\end{equation}
The subscripts ``$k$'' here are a reminder that these two quantities are fixed
once the total energy-momentum and spectator momentum are specified.
The only remaining degree of freedom is the direction of the momenta of one of the
non-spectator pair in the two-particle CM frame, which we denote $\hat a^*$. 
In summary, with $E$ and $\vec P$ fixed, 
the configuration of three on-shell particles is specified by $(\vec k, \hat a^*)$. 
It is also useful to decompose the dependence on $\hat a^*$ into
spherical harmonics, e.g.
\begin{equation}
\overline\Gamma(\vec k, \hat a^*) = 
\sqrt{4 \pi} \sum_{\ell m}  Y_{\ell m}(\hat a^*) \overline\Gamma_{\ell m}(\vec k)  \,.
\end{equation}

Up to this stage, the bound state has an unspecified total angular momentum.
We now follow MRR and make two assumptions: first that s-wave two-particle channels
dominate, and second that the total angular momentum of the bound state is zero.
Given this, we can set $\ell=m=0$, so that only $\overline\Gamma_{00}$ contributes.
In addition, we note that $\overline\Gamma_{00}(\vec k)$ and $\Gamma_{00}(\vec p)$ cannot depend on the directions of $\vec k$ and $\vec p$, since, for each function, there is no other direction defined in the CM frame.
We then use the abbreviation 
$\overline\Gamma(k) \equiv \overline\Gamma_{00}(\vec k)$,
and similarly for $\Gamma(p)$. 
Thus the pole form (\ref{eq:pole}) becomes
\begin{equation}
\label{eq:poleS}
\mathcal M_3(\vec p; \vec k) 
\sim - \frac{\Gamma( p ) 
\overline \Gamma(  k)}{E^{*2} - E_B^{2} } \,.
\end{equation}

We now confine this system to a finite spatial cube with side-length $L$ and apply 
periodic boundary conditions. 
The infinite-volume bound-state energy is then shifted to a FV value given by
\begin{equation}
E_B(L) = E_B + \Delta E(L) \,. 
\end{equation}
We are interested in the large $L$ regime,
\begin{equation}
\frac{1}{L} \ll \kappa \ll m \,,
\label{eq:hier}
\end{equation}
in which the energy shift $\Delta E(L)$ is much smaller in magnitude
than the binding energy $\kappa^2/m$. 
Our aim is to find the leading dependence on $L$ as $L\to\infty$.
 We stress that this is probably not a practical limit
[since achieving the
hierarchy of Eq.~(\ref{eq:hier}) requires very large boxes]
but since we control the $L$-dependence analytically we can consider
arbitrarily large values.

MRR determine the energy shift using NRQM in the unitary limit, assuming only two-particle
potentials (i.e. including no three-particle potential), 
and also assuming that s-wave scattering dominates.
They find 
\begin{equation}
\label{eq:result}
\Delta E(L) = c \vert A \vert^2 \frac{\kappa^2}{m} 
\frac{1}{(\kappa L)^{3/2}} e^{- 2 \kappa L/\sqrt{3} }+ \cdots \,,
\end{equation}
where the ellipsis indicates terms suppressed by additional powers of 
$\kappa/m$ or $1/(\kappa L)$, as well as subleading exponentials. 
We will use the ellipsis in this fashion henceforth.

The numerical coefficient $c$ in Eq.~(\ref{eq:result}) was determined by MRR from the
solution to the Faddeev equation. It is
\begin{multline}
c = - \frac92 \cdot 3^{3/4} \sqrt{ \pi}    \sh(\pi s_0) \sh^2\!\left(\frac{\pi s_0}{2}\right)
\\ 
\times
\left(  \frac34 \sh(\pi s_0) - \frac{3 \pi s_0}{4} 
      - \frac{4 \pi}{\sqrt{3}} \sh \frac{\pi s_0}{3} 
      + \frac{2 \pi}{\sqrt{3}} \sh \frac{2 \pi s_0}{3}  \right )^{-1} \,,
\label{eq:cdef}
\end{multline}
where ``$\sh$'' is an abbreviation for ``$\sinh$'' and $s_0$ is the solution to
\begin{equation}
s_0 \cosh \frac{\pi s_0}{2} = \frac{8}{\sqrt 3} \sinh \frac{\pi s_0}{6} \,.
\label{eq:s0def}
\end{equation}
The numerical values are
$s_0  \simeq 1.00624 $ %\, 237\, 825 
and
$c  \simeq -96.351$. % \,838\,953 

The factor $\vert A \vert^2$ in Eq.~(\ref{eq:result}) is a normalization coefficient. 
It arises because the three-body wavefunction used by MRR to derive their result
(and also used indirectly in the present article) is
not strictly a solution to the Schr\"odinger equation. It has the
correct asymptotic form when the three particles are well separated,
but fails at short distances. Nonetheless, the approximate
wavefunction gives the correct leading prediction for $\Delta E(L)$,
as long as one accounts for a possible normalization discrepancy. If
$\psi_{\mathrm{true}}$ is the true wavefunction and
$\psi_{\mathrm{asymp}}$ is the approximation used here (both
normalized), then $A$ is defined by 
$\psi_{\mathrm{true}}/\psi_{\mathrm{asymp}} \longrightarrow A$ where the arrow indicates the
limit of all particles being well separated. In other words 
$A \psi_{\mathrm{asymp}}$ (and not just $\psi_{\mathrm{asymp}}$) is the
wavefunction that correctly predicts the the energy shift. In the
present study, the approximate wave function is needed to determine
the value of the residue factors $\Gamma$ and $\overline \Gamma$ in
the unitary theory. Thus the coefficient $A$ enters our prediction for
the energy shift through these quantities.

It is interesting to compare Eq.~(\ref{eq:result}) to the corresponding result
for two particles~\cite{Beane:2003da}
\begin{equation}
\Delta E_2(L) = - 12 \frac{\kappa_2^2}{m} \frac1{\kappa_2 L} e^{-\kappa_2 L} + \cdots
\,.
\label{eq:result2}
\end{equation}
Here we have assumed that the system is near the unitary limit, with a large
positive scattering length so that there is a bound state.
The binding momentum $\kappa_2$ is defined so that
the binding energy is 
\begin{equation}
E_{B_2}=2m-\frac{\kappa_2^2}{m} \,.
\label{eq:EB2}
\end{equation}
The result (\ref{eq:result2}) follows directly from 
L\"uscher's quantization condition~\cite{Luscher:1991cf}
(as we review in Sec.~\ref{sec:twopart}).
We see that it differs from the three-particle result not only in the exponent and
power of $L$, but also in having a much simpler overall constant.
We return to the comparison between two- and three-particle results 
in Secs.~\ref{sec:twopart} and Sec.~\ref{sec:moving}.

\section{Determining the energy shift from the quantization condition}

\label{sec:QCexp}

In this section we demonstrate that our quantization condition leads
to the following prediction for the energy shift in the unitary limit
\begin{equation}
\label{eq:DE}
\Delta E(L) = - \frac{1}{2 E_B} 
\bigg[ \frac1{L^3} \sum_{\vec k} - \int_{\vec k} \bigg]  
\frac{\overline \Gamma^{(u)} \!(k) \,  \Gamma^{(u)}(k)  }{2 \omega_k \mathcal M_2^s(k)} 
+\cdots \,.
\end{equation}
Here $\vec k$ is the spectator momentum described in the previous section
(with $k$ its magnitude). It is summed over all integer three-vectors multiplied
by $2\pi/L$, whereas for the integral we use the shorthand
$\int_{\vec k} \equiv \int d^3 k/(2 \pi)^3$. 
$\mathcal M_{2}^s(k)$ is the two-to-two s-wave scattering amplitude,
evaluated at the CM energy $E_{2,k}^*$ of the nonspectator pair.
Since this energy depends on $k$, we follow the notation of Ref.~\cite{SpectoK} and 
denote this explicitly.
 $\mathcal M_{2}^s(k)$ also depends on the total energy $E$, 
which is here set to the bound-state energy $E_B$.
The residue factors $\Gamma^{(u)}$ and $\overline\Gamma^{(u)}$ are those defined in
in Eqs.~(\ref{eq:Gammadef}) and (\ref{eq:Gammabardef}),
except that they are projected to the s-wave and {\em unsymmetrized}.
We define them precisely in Eq.~(\ref{eq:poleuu}) below.

\bigskip
To set up the derivation of Eq.~(\ref{eq:DE}) 
we need to recall some details of the three-particle quantization condition.
It turns out to be convenient to
reformulate the quantization condition of Ref.~\cite{SpectoK}
using the developments of Ref.~\cite{KtoM}.\footnote{%
This is the same reformulation that simplifies the development of the threshold
expansion~\cite{thresh}.}
Thus, rather than consider a general three-particle correlator as in Ref.~\cite{SpectoK},
we focus on the quantity $\MthL$, which is defined in Ref.~\cite{KtoM}
and referred to there as the ``finite-volume three-particle scattering amplitude''.
From the point of view of the quantization condition, $\MthL$ is just a particular
three-particle correlator in finite spatial volume,\footnote{%
The time direction has infinite extent.
}
so the positions of its poles determine
the FV spectrum. It indeed leads to the same quantization condition as given 
in Ref.~\cite{SpectoK}.
Its advantage here is that it goes over to the standard infinite-volume
scattering amplitude, $\Mth$, when $L\to\infty$, which allows us to make contact with the
known pole-form (\ref{eq:poleS}) of the 
standard scattering amplitude.\footnote{%
The $L\to\infty$ limit must be taken in a particular way with an $i\epsilon$ prescription
as explained in Ref.~\cite{KtoM}.}

Combining Eqs.~(39) and (68) of Ref.~\cite{KtoM} we have
\begin{equation}
\label{eq:M3L}
\mathcal M_{3,L} = \mathcal S \Big [ \mathcal M^{(u,u)}_{3,L} \Big ] \,,
\end{equation}
where $\mathcal S$ is the symmetrization operator, 
and
\begin{equation}
\label{eq:M3Luu}
  \mathcal M^{(u,u)}_{3,L} \equiv \mathcal D^{(u,u)}_L + 
\mathcal L^{(u)}_L \frac{1}{1 + \Kdf F_3} \Kdf \mathcal R^{(u)}_L  \,,
\end{equation}
is the unsymmetrized finite-volume scattering amplitude.
The form of the quantization condition used here is that the FV spectrum
is given by the poles in $\MthL^{(u,u)}$.
These occur at energies such that $\det(1 + \Kdf F_3)=0$, which is the original
form of the quantization given in Ref.~\cite{SpectoK}.

Equation~(\ref{eq:M3Luu}) is written in a compact notation that we now explain.
First, we note that all quantities are matrices in the space of
discrete spectator momenta.\footnote{%
Strictly speaking, this holds only for the ``internal'' momenta that are summed
over in matrix products. External momenta can take any values. This subtlety is
discussed in Ref.~\cite{KtoM}.}
For example $F_3 = F_{3;k' k}$ where $k'$ and $k$ are shorthand for
$\vec k', \vec k \in (2 \pi/L) \mathbb Z^3$. 
Generally these matrices also have two sets of angular momentum indices, 
but these are absent in the present case, 
since we only include the s-wave component of both two and three-particle 
scattering quantities. This approximation mirrors that made by MRR.

We next explain the unsymmetrized nature of the
quantities in Eq.~(\ref{eq:M3Luu}), indicated by the superscript $(u)$.
This lack of symmetrization is defined in the context of a diagrammatic description
of $\mathcal M^{(u,u)}_{3,L}$. For diagrams that involve a two-to-two vertex 
next to the external legs of either the initial or final state, the $(u)$ indicates that
this insertion always scatters the two particles with total momentum $- \vec k$ 
(the non-spectator pair).
$\mathcal M^{(u,u)}_{3,L}$ and $\mathcal D^{(u,u)}_L$ have two superscripts because
this rule applies to both initial and final momenta, while
${\cal L}_L^{(u)}$ and ${\cal R}_L^{(u)}$ have one each since they involve,
respectively, only the final and initial momenta.
Thus the indices $k'$ and $k$
denote the momenta of the particles that are unscattered by
the outermost two-to-two vertices. 
The operator $\mathcal S$ symmetrizes the
momenta by setting the initial state momentum index to the three possible values 
$ \{\vec k, \vec a, - \vec k - \vec a \}$, and the final state index to the 
corresponding three choices, and then summing the resulting nine terms.
For further details see Refs.~\cite{SpectoK,KtoM}. 

We now turn to $\Kdf$, the divergence-free three-particle K matrix.
This is the only quantity in Eq.~(\ref{eq:M3Luu}) that (for fixed
external indices) has no volume dependence. We define this modified K
matrix in Ref.~\cite{SpectoK} and give its relation to the standard
three-to-three scattering amplitude in Ref.~\cite{KtoM}. Indeed, the
relation between $\Kdf$ and the scattering amplitude, $\Mth$, is
derived in Ref.~\cite{KtoM} by first proving Eq.~(\ref{eq:M3L}) and
then taking a careful infinite-volume limit. 
Within our formalism, $\Kdf$ plays the role of an effective, quasilocal
three-particle interaction. It is, however, not a physical quantity as
it depends on a cutoff function ($H$, to be described shortly).

Finally, we give the explicit forms for the various finite-volume matrices 
appearing in Eq.~(\ref{eq:M3Luu}).
We begin with the part of $\MthL^{(u,u)}$ that involves only two-particle
interactions:
\begin{equation}
\label{eq:DLdef}
\mathcal D^{(u,u)}_L \equiv -\frac{1}{1 + \ML G} \ML G [2 \omega L^3] \ML \,,
\end{equation}
where 
\begin{align}
\label{eq:omegamatdef}
\left[ {2 \omega L^3} \right]_{k' k} & \equiv
\delta_{k'k}   {2 \omega_k L^3}\,.
\end{align}
$\ML$ is the two-particle finite-volume scattering amplitude
\begin{align}
\label{eq:MLdef}
\ML & \equiv \mathcal M^s_2 \frac{1}{1 + F^{i \epsilon} \mathcal M^s_2 } \,, 
\end{align}
with the matrix form of the s-wave scattering amplitude being
\begin{equation}
\mathcal M^s_{2;k'k} = \delta_{k'k} \mathcal M^s_2(k)\,,
\end{equation}
while $F^{i\epsilon}$ is the moving-frame s-wave L\"uscher zeta function,
\begin{align}
\begin{split}
F^{i\epsilon}_{k' k}
&=
\delta_{k'k} \frac12 \bigg[\frac{1}{L^3} \sum_{\vec p} - \int_{\vec p} \bigg]  \\[-3pt] & \times
\frac{   H(\vec k)
  H(\vec p\,)H(\vec b_{kp})}
{2 \omega_p 2 \omega_{kp}(E - \omega_k - \omega_p -  \omega_{kp} + i \epsilon)}\,,
  \end{split}
\end{align}
with
\begin{gather}
\omega_{kp}  = \sqrt{ b_{kp}^2 + m^2} \,, \ \ \ \ \
   \vec b_{kp}  = - \vec p - \vec k  \,. % \\
\end{gather}
$H(\vec k)$ is a smooth cutoff function that vanishes when $k$ becomes
large enough [of ${\cal O}(m)$] that the nonspectator pair has an energy $E_{2,k}^{*2}\le 0$.
The precise form of $H$, given in Ref.~\cite{SpectoK}, will not be needed here.
The propagator $G$ is
\begin{align}
\label{eq:Gdef}
G_{p  k  } 
& \equiv
\frac{ H(\vec p\,) H(\vec k\,)} 
{2 \omega_{kp} (E - \omega_k - \omega_p - \omega_{kp})}
\frac{1}{2 \omega_k L^3} \,.
\end{align}
This arises from the parts of Feynman diagrams where
three particles propagate between two-to-two vertices in which the
scattering pair changes.

The quantities in the second term in Eq.~(\ref{eq:M3Luu}) are
\begin{align}
\mathcal L^{(u)}_L  & =   \frac{1}{3} - \frac{1}{1 + \ML G} \ML F   \,, 
\label{eq:LLudef}\\
\mathcal R^{(u)}_L  & =    \frac{1}{3} - \frac{F}{2 \omega L^3} 
\frac{1}{1 + \ML G} [2 \omega L^3] \ML    
\label{eq:RLudef}\,, \\
F_3 & = \frac{F}{2 \omega L^3}  \mathcal L_L^{(u)} \,,
\label{eq:F3def}
\end{align}
where $F$ differs from $F^{i\epsilon}$ by a phase-space term:
\begin{align}
\label{eq:Fdef3}
F_{k' k} & = F^{i\epsilon}_{k' k} + \rho_{k' k} \,, 
\\
  \rho_{k'k} & \equiv \delta_{k'k} H(\vec k) \widetilde \rho (E_{2,k}^*) \,,
\\
  \begin{split}
\widetilde\rho(P_2) &\equiv \frac{1}{16 \pi  \sqrt{P_2^2}}  \\[-3pt] & \times
\begin{cases} 
-  i \sqrt{P_2^2/4-m^2} & (2m)^2< P_2^2 \,, 
\\ 
\vert \sqrt{P_2^2/4-m^2} \vert &   0< P_2^2 \leq (2m)^2 \,.
\end{cases}
\label{eq:rhodef2}
\end{split}
\end{align}

We stress again that all of these quantities have been projected to the 
s-wave component of the spectator pair angular dependence.
We can now see more clearly why this is appropriate, given that we are
matching to the MRR calculation in which there are only s-wave two-particle potentials.
Begin by considering ${\cal D}_L^{(u,u)}$,  defined in Eq.~(\ref{eq:DLdef}),
 which is the first term in $\MthL^{(u,u)}$, itself defined in Eq.~(\ref{eq:M3Luu}).
Expanding the denominator in a geometric series, we find a sum
of terms each containing alternating factors of
the two-particle finite-volume scattering amplitude,
$\mathcal M_{2,L}$, and the three-particle propagator $G$.
If the two-particle scattering is dominated by the s-wave channel,
i.e if $\M \approx \M^s$,
then $\ML$ is also pure s-wave,\footnote{%
This follows from the general form $\ML=\M - \M F^{i\epsilon} \M + \cdots$,
which holds for arbitrary angular momenta with the general form for $F^{i\epsilon}$.}
and this projects the $G$ factors onto their s-wave components.
This projection has already been included in the equations above.

Next we turn to the second term in $\MthL^{(u,u)}$.
Here the same projection onto s-wave components works for $G$.
For $\Kdf$, however, there is no such projection, due to the 
factors of $1/3$ in ${\cal L}_L^{(u)}$ and ${\cal R}_L^{(u)}$.
Thus we must assume that $\Kdf$ itself contains only s-wave nonspectator-pair
components. 
This is reasonable as the MRR calculation has no three-body potential.
Naively, one might think that this would imply that we could set $\Kdf=0$,
in which case the second term in Eq.~(\ref{eq:M3Luu}) would simply vanish.
However, since ${\cal D}_L^{(u,u)}$ depends on the cutoff $H$, it is not physical
by itself. The $\Kdf$ term is needed to cancel its high-momentum cutoff dependence.
$\Kdf$ is thus a short-distance quantity, and an s-wave approximation is reasonable.

\medskip
We now return to the aim of this section: using Eq.~(\ref{eq:M3Luu}) to derive the
result (\ref{eq:DE}). Our approach is simply to pull out the leading $L$-dependence
contained in $\mathcal D_L^{(u)}$, $\mathcal L_L^{(u)}$, 
$\mathcal R_L^{(u)}$ and $F_3$, and 
then make use of the fact that
$\MthL^{(u,u)}$ and $\Mth$ have nearby poles.
It is pedagogically simpler to proceed in two stages, first setting $\Kdf=0$
and then considering the general case.

\subsection{Analysis for $\Kdf=0$}
\label{subsec:Kzero}

As noted above, the choice $\Kdf=0$ can only approximately correspond to the MRR
calculation. It implies that $\MthL^{(u,u)}={\cal D}_L^{(u,u)}$
and $\Mth={\cal S}\big[\mathcal D_L^{(u,u)}\big]$, so that
$\Mth$ depends on the cutoff function $H$. 
Nevertheless, it is instructive to first consider this case to
see how Eq.~(\ref{eq:M3Luu}) arises in a simpler context.

As can be seen from Eq.~(\ref{eq:DLdef}), $L$-dependence
enters ${\cal D}_L^{(u,u)}$ through the factors of $F^{i\epsilon}$ contained in $\ML$,
and through the presence of momentum sums (rather than integrals) in the
matrix products, as well as the explicit factor of $2\omega L^3$. 
It turns out, as we show in Appendix \ref{app:approx}, 
that the $F^{i\epsilon}$ contributions are subdominant,
suppressed by a factor of $1/(\kappa L)$. 
Thus to obtain the leading volume dependence we can set $F^{i \epsilon}=0$,
implying [via Eq.(\ref{eq:MLdef})] that we can replace $\ML$ with $\M^s$.
Dropping the superscript $s$ to reduce notational clutter, we thus have
\begin{equation}
\label{eq:DLred1}
\mathcal D^{(u,u)}_L = -\frac{1}{1 + \M G} \M G [2 \omega L^3] \M + \cdots\,.
\end{equation}

To pull out the FV dependence, we must first understand the $L\to\infty$ limit
of $\mathcal D^{(u,u)}_L$, which we call $\mathcal D^{(u,u)}$.
In general, this limit must be taken carefully, since the summands
(and, in particular, the factors of $G$) 
have poles that require a prescription when the sums become integrals.
As explained in Ref.~\cite{KtoM}, the appropriate choice
is to first shift the poles in $G$ by $i \epsilon$, 
then send $L \to \infty$, and finally take $\epsilon \to 0$.\footnote{%
Strictly speaking, when, as here, we consider below-threshold energies, the sums and integrals
never run over the poles, so no $i\epsilon$ is needed. 
Nevertheless, we include these factors so that the results hold also above threshold.}
This is the choice of limits that sends $\ML$ to $\M$ and $\MthL$ to $\Mth$.
Thus we have\footnote{%
This equation defines $\mathcal D^{(u,u)}$ also if all factors of $\M$ are replaced
by $\ML$, i.e. if the full $\mathcal D_L^{(u,u)}$ is used. Similarly, 
Eq.~(\ref{eq:Duuint}) remains true in the general case.}
\begin{multline}
\mathcal D^{(u,u)}(\vec k, \vec p) \\ 
\equiv - \lim_{\epsilon \to 0} \lim_{L \to \infty} 
\left[  \frac{1}{1 + \M G} \M G [2 \omega L^3] \M \right]_{kp} \,,
\end{multline}
a quantity already introduced in Eq.~(85) of Ref.~\cite{KtoM}. 
It satisfies the integral equation
\begin{multline}
\label{eq:Duuint}
\mathcal D^{(u,u)}(\vec k, \vec p) = 
- \mathcal M(\vec k) G^\infty(\vec k, \vec p) \mathcal M(\vec p) 
\\ 
- \int_{\vec \ell} \frac{1}{2 \omega_{\ell}} \mathcal M(\vec k) 
G^\infty(\vec k, \vec \ell) \mathcal D^{(u,u)}(\vec \ell, \vec p)\,,
\end{multline}
where the infinite-volume propagator is
\begin{equation}
G^\infty(\vec p, \vec k) \equiv \frac{H(\vec p) H(\vec k)}{2 \omega_{kp} (E - \omega_k - \omega_p - \omega_{kp} + i \epsilon)}  \,.
\end{equation}

Our aim now is to express the finite-volume matrix $\mathcal D^{(u,u)}_L$ 
in terms of the infinite-volume function $\mathcal D^{(u,u)}$.
To do so we expand the former in powers of $\M$
\begin{align}
\mathcal D^{(u,u)}_L & = \sum_{n=2}^\infty \mathcal D^{(n,u,u)}_L \,, \\
\mathcal D^{(n,u,u)}_L & \equiv  -   [- \M G]^{n-2} \M G [2 \omega L^3] \M \,.
\end{align}
Defining $\mathcal D^{(n,u,u)}$ to be the infinite-volume limit
of $\mathcal D^{(n,u,u)}_L$,
we also have
\begin{align}
\mathcal D^{(u,u)} & = \sum_{n=2}^\infty \mathcal D^{(n,u,u)} \,.
\end{align}
We now can relate $\mathcal D_L^{(n,u,u)}$ to $\mathcal D^{(n,u,u)}$ order by order.
The lowest order is simple:
\begin{equation}
\mathcal D^{(2,u,u)}_{L,kp} = \mathcal D^{(2,u,u)}(\vec k, \vec p) 
= - \mathcal M(\vec k) G^{\infty}(\vec k, \vec p) \mathcal M(\vec p) \,. 
\end{equation}
For $n=3$ we find
\begin{align}
\label{eq:D3Ldecomstart}
\mathcal D^{(3,u,u)}_L & =   \M G   \M G [2 \omega L^3] \M  \,, \\
&  \hspace{-20pt} =   \M G [2 \omega L^3]   \M  \frac{1}{2 \omega L^3 \M} \M G [2 \omega L^3] \M \,, \\[5pt]
\begin{split}
&  \hspace{-20pt} = \mathcal D^{(3,u,u)}(\vec k, \vec p) \\
& \hspace{-10pt} +  \bigg [ \frac{1}{L^3} \sum_{\vec \ell} - \int_{\vec \ell} \bigg ]              \frac{   \mathcal D^{(2,u,u)}(\vec k, \vec \ell) \mathcal D^{(2,u,u)}(\vec \ell, \vec p)  }{2 \omega_{\ell} \M( \ell)}  \,.
\label{eq:D3Ldecom}
\end{split}
\end{align}
In the second line we multiplied and divided by $[2 \omega L^3] \M$ and 
in the third we expressed the implicit sum as an integral plus a sum-integral difference.

To continue this pattern to higher orders it is helpful to use a compact notation
in which we write Eq.~(\ref{eq:D3Ldecom}) as
\begin{equation}
\label{eq:D3Ldecomshort}
\mathcal D^{(3,u,u)}_L  = \mathcal D^{(3)} 
+ \mathcal D^{(2 )} \mathcal C^{(-1)}  \mathcal D^{(2 )}  \,.
\end{equation}
Here we have dropped the $u$ superscripts on the right-hand side 
and have introduced $\mathcal C^{(-1)}$ to represent the sum-integral difference ``cut''. 
Its precise definition can be inferred by comparing Eqs.~(\ref{eq:D3Ldecom}) 
and (\ref{eq:D3Ldecomshort}). 
The $(-1)$ superscript indicates that the cut has one factor of $\M$ in the denominator. 
Since $\mathcal D^{(n,u,u)}_L$ is defined by the number of $\M$ insertions, 
it is convenient to track these in the decomposition on the right-hand side. 
In particular, the superscripts of the second term must sum to three. 
This pattern persists to all orders, so that the decomposition 
of $\mathcal D^{(n,u,u)}_L$ may be defined as the sum of all possible 
terms built from alternating factors of $\mathcal D^{(m)}$ and $\mathcal C^{(-1)}$ 
whose superscripts sum to $n$. 
For example, the $n=4$ result is
\begin{multline}
\mathcal D^{(4,u,u)}_L  = \mathcal D^{(4)} 
+ \mathcal D^{(3 )} \mathcal C^{(-1)}  \mathcal D^{(2 )}  
\\ 
+  \mathcal D^{(2 )} \mathcal C^{(-1)}  \mathcal D^{(3 )}  
+  \mathcal D^{(2)} \mathcal C^{(-1)}   \mathcal D^{(2)} \mathcal C^{(-1)}  \mathcal D^{(2 )} 
\,.
\end{multline}

Continuing in this fashion to all orders and summing the result gives
\begin{equation}
\mathcal D_L^{(u,u)} = \sum_{n=0}^\infty \mathcal D^{(u,u)} \left [ \mathcal C^{(-1)} \mathcal D^{(u,u)} \right ]^n + \cdots \,.
 \label{eq:DLseries}
\end{equation}
This can be succinctly written as an integral equation
\begin{multline}
\label{eq:DLdecomposed}
\mathcal D_L^{(u,u)}(\vec k, \vec p)  = \mathcal D^{(u,u)}(\vec k, \vec p) \\
 +  \bigg [ \frac{1}{L^3} \sum_{\vec \ell} - \int_{\vec \ell} \bigg ]  
\mathcal D^{(u,u)}(\vec k, \vec \ell)  
     \frac{1}{2 \omega_{\ell} \M( \ell)}     \mathcal D_L^{(u,u)}(\vec \ell, \vec p) + \cdots \,.
\end{multline}
Here we have extended the definition of $\mathcal D_L^{(u,u)}(\vec k, \vec p)$ 
to continuous values of $\vec k$ and $\vec p$. 
This extension is straightforward given the definitions of the building blocks in 
Eqs.~(\ref{eq:omegamatdef})-(\ref{eq:rhodef2}) 
as was already discussed in Ref.~\cite{KtoM}.  

Now we observe that, since (with $\Kdf=0$) $\Mth = \mathcal S [ \mathcal D^{(u,u)}]$,
and recalling that $\Mth$ has the pole form (\ref{eq:poleS}),
then $\mathcal D^{(u,u)}$ itself must have a pole associated with the bound state. 
Symmetrization cannot lead to the development of a pole.
We parametrize the pole in $\mathcal D^{(u,u)}$ as
\begin{equation}
\label{eq:Dpole}
\mathcal D^{(u,u)}(\vec k', \vec k)\bigg|_{\Kdf=0} \sim - 
\frac{ \Gamma^{(u)}(k')  \, \overline \Gamma^{(u)} \!(k) }{E^2 - E_B^2} \,,
\end{equation}
where we have added an explicit reminder
that we are working in the $\Kdf=0$ approximation.
We also know that $\MthL^{(u,u)}=\mathcal D_L^{(u,u)}$ must have a nearby pole, 
corresponding to the bound state with its energy slightly shifted:
\begin{equation}
\label{eq:DLpole}
\mathcal D_L^{(u,u)}(\vec k', \vec k)\bigg|_{\Kdf=0} \sim - \frac{ \Gamma_L^{(u)}(k')  
\, \overline \Gamma_L^{(u)} \!(k) }{E^2 - (E_B + \Delta E(L))^2} \,.
\end{equation}
Here we have also allowed for a finite-volume dependence in the residue factors.

Substituting Eqs.~(\ref{eq:Dpole}) and (\ref{eq:DLpole}) into
Eq.~(\ref{eq:DLdecomposed}), multiplying by both poles and dividing by the
common residue factors, we find that the residue factors are volume independent 
\begin{align}
\Gamma_L^{(u)}(k') \overline \Gamma_L^{(u)}(k) &=
\Gamma^{(u)}(k') \overline \Gamma^{(u)}(k) + \cdots \,,
\label{eq:residuesmatch}
\end{align}
and that
\begin{multline}
-   [E^2 - E_B^2]  =-  [E^2 - (E_B + \Delta E(L))^2]  
 \\
 +  \bigg [ \frac{1}{L^3} \sum_{\vec \ell} - \int_{\vec \ell} \bigg ]      
\frac{ \overline \Gamma^{(u)} \!(\ell) \   \Gamma^{(u)}(\ell)  }{2 \omega_{\ell} \M( \ell)}  + \cdots
\,.
\label{eq:almostDE}
\end{multline}
Both of these results have corrections that are of higher order in $\Delta E(L)$
if one accounts for the finite residues beneath the poles in Eqs.~(\ref{eq:Dpole})
and (\ref{eq:DLpole}). In particular, there are $\mathcal O(\Delta E^2)$ corrections to
Eq.~(\ref{eq:almostDE}). However, these are suppressed relative to the leading terms
and thus can be dropped in our calculation.
The final step is to solve Eq.~(\ref{eq:almostDE}) for $\Delta E(L)$,
which, after and dropping terms of $\mathcal O(\Delta E^2)$, 
leads to the desired result, Eq.~(\ref{eq:DE}). 

We emphasize that all of the approximations leading to Eq.~(\ref{eq:DE}) are justified by our aim to only determine the leading finite-volume shift of the three-particle bound state. In particular, all neglected terms lead to contributions to $\Delta E(L)$ that vanish faster than the term we are after in the large $L$ limit. This is motivated not only by the aim to approximate $\Delta E(L)$ for large volumes, but also because we are pursuing the same expansion as MRR in order to reproduce their result.

\subsection{Analysis for general $\Kdf$}
\label{subsec:Knonzero}

We now extend the analysis to nonzero $\Kdf$, requiring that we keep the
second, $\Kdf$-dependent term in Eq.~(\ref{eq:M3Luu}). 
As for $\Kdf=0$, we argue in Appendix \ref{app:approx} that
finite-volume effects from factors of $F^{i\epsilon}$ are subleading compared
to those we keep. This considerably simplifies the analysis.

We expand the second term in Eq.~(\ref{eq:M3Luu}) in powers of $\Kdf$ and
focus first on the contribution containing a single factor of $\Kdf$ 
\begin{equation}
\MthL^{(u,u)} \supset   \mathcal L^{(u)}_L \Kdf \mathcal R_L^{(u)}   \,.
\end{equation}
Setting $F^{i\epsilon}=0$ (so that $F \to \rho$) we have
\begin{align}
\mathcal L^{(u)}_L  & =   \frac{1}{3} - \frac{1}{1 + \M G} \M \rho  + \cdots  \,, \\
\mathcal R^{(u)}_L  & =    \frac{1}{3} - \frac{\rho}{2 \omega L^3} \frac{1}{1 + \M G} [2 \omega L^3] \M   + \cdots \,. 
\end{align}
We expand in powers of $G$ to reach
\begin{equation}
\label{eq:sumLsumR}
 \mathcal L^{(u)}_L \Kdf \mathcal R_L^{(u)} = 
\sum_{m,n=1}^\infty   \mathcal L^{(m,u)}_L \Kdf \mathcal R_L^{(n,u)} \,,
\end{equation}
where
\begin{align}
\mathcal L^{(0,u)}_L  & =   \tfrac{1}{3} - \M \rho \,, \\
\mathcal L^{(n,u)}_L  & =  - [- \M G]^{n} \M \rho \qquad (n>1)\,,
\end{align}
and similarly for $\mathcal R^{(n,L)}$. 

For the terms with one or more factors of $G$, 
the intermediate sums from contracted indices are now decomposed into integrals and
sum-integral differences as in Eqs.~(\ref{eq:D3Ldecomstart})-(\ref{eq:D3Ldecom}) above. 
For example
\begin{align}
\mathcal L^{(1,u)}_L \Kdf  & = \M G \M \rho \Kdf \,, \\
& \hspace{-30pt} = [- \M G [2 \omega L^3] \M] \frac{1}{2 \omega L^3 \M} [- \M \rho \Kdf] \,, \\
\begin{split}
& \hspace{-30pt} = \mathcal L^{(1,u)} \Kdf \\
& \hspace{-50pt} + \bigg [ \frac{1}{L^3} \sum_{\vec \ell} - \int_{\vec \ell}   \bigg ]
  \frac{\mathcal D^{(2,u,u)}(\vec p, \vec \ell ) \ \mathcal L^{(0/\rho,u)} \Kdf
  (\vec \ell, \vec k) }{2 \omega_\ell L^3 \M (  \ell)} \,.
\end{split}
\label{eq:L1uK}
\end{align}
Here $\mathcal L^{(1,u)}$ is understood as an integral operator
\begin{multline}
\mathcal L^{(1,u)} \Kdf \equiv \\ \M(\vec p) \int_{\vec \ell} \frac{1}{2 \omega_{\ell}}
 G(\vec p, \vec \ell) \M(\vec \ell) \rho(\vec \ell) \Kdf(\vec \ell, \vec k) \,.
\end{multline}
We have also introduced 
\begin{equation}
\mathcal L^{(0/\rho,u)} \equiv - \M \rho \equiv \mathcal L^{(0,u)} - \tfrac13\,.
\label{eq:L0rho}
\end{equation}

Switching to the shorthand introduced in Eq.~(\ref{eq:D3Ldecomshort})  we 
rewrite Eq.~(\ref{eq:L1uK}) as the action of
\begin{equation}
\mathcal L^{(1,u)}_L = 
\mathcal L^{(1 )} + \mathcal D^{(2 )} \mathcal C^{(-1)} \mathcal L^{(0/\rho )}
\end{equation}
on $\Kdf$.
The next order is given by
\begin{multline}
\mathcal L^{(2,u)}_L = \mathcal L^{(2 )} 
+ \mathcal D^{(3 )} \mathcal C^{(-1)} \mathcal L^{(0/\rho )}\\ 
 + \mathcal D^{(2 )} \mathcal C^{(-1)} \mathcal L^{(1 )}  
+ \mathcal D^{(2 )} \mathcal C^{(-1)} \mathcal D^{(2 )} 
\mathcal C^{(-1)} \mathcal L^{(0/\rho )}\,.
\end{multline}
The pattern generalizes as in the previous subsection:
the $n$th-order result is the sum of all terms built 
from alternating factors of $\mathcal D$ and $\mathcal C^{(-1)}$, followed by a factor
of $\mathcal L$,
subject to the condition that the superscripts sum to $n$. 
Repeating the exercise for $\mathcal R_L^{(u)}$,
substituting into Eq.~(\ref{eq:sumLsumR}), and summing over powers of $G$, 
we find
\begin{multline}
\mathcal L^{(u)}_L \Kdf \mathcal R_L^{(u)} = 
\bigg [ \mathcal L^{(u)}  + \sum_{n=1}^\infty 
\left[ \mathcal D^{(u,u)} \mathcal C^{(-1)} \right ]^n   
(\mathcal L^{(u)}-\tfrac13)    \bigg ] 
\\ 
\times \Kdf  \bigg [ \mathcal R^{(u)}  + (\mathcal R^{(u)}-\tfrac13) \sum_{n=1}^\infty   
 \left[ \mathcal C^{(-1)} \mathcal D^{(u,u)}  \right ]^n  \bigg ]  \,.
\label{eq:LKR}
\end{multline}
Here ${\cal L}^{(u)} = \sum_{n=0}^\infty {\cal L}^{(n,u)}$ is
the infinite-volume limit of ${\cal L}_L^{(u)}$, and similarly for 
${\cal R}^{(u)}$.\footnote{%
Explicit forms for ${\cal L}^{(u)}$ and ${\cal R}^{(u)}$
are given in Eqs.~(92) and (94) of Ref.~\cite{KtoM}, respectively.
In that work these quantities are, however, denoted with a double superscript,
e.g. ${\cal L}^{(u,u)}$.}

The factors of $(-1/3)$ in Eq.~(\ref{eq:LKR}) arise because of the difference between 
$\mathcal L_L^{(0,u)}$ and $\mathcal L^{(0/\rho,u)}$---see Eq.~(\ref{eq:L0rho}).
It turns out, however, that these factors lead to contributions with
subleading dependence on $L$, as explained in Appendix \ref{app:approx}.
Thus we can drop them and obtain
\begin{multline}
\mathcal L^{(u)}_L \Kdf \mathcal R_L^{(u)} = 
\left( \sum_{n=0}^\infty \left[ \mathcal D^{(u,u)} \mathcal C^{(-1)} \right ]^n  \right)
\mathcal L^{(u)}    
\\ \times \Kdf   \sum_{n=0}^\infty   \mathcal R^{(u)}  \left(
\left[ \mathcal C^{(-1)} \mathcal D^{(u,u)}  \right ]^n  \right)  + \cdots \,.
\label{eq:LLKRLdecomp}
\end{multline}
We observe that the leading volume dependence comes from an alternating
series of factors of $\mathcal D^{(u,u)}$ and $\mathcal C^{(-1)}$ that appear only
on the ends of the expression. This is the same series that appears in the
expression for $\mathcal D_L^{(u,u)}$, Eq.~(\ref{eq:DLseries}).

To complete the pattern we need to study the next contribution to the second term in 
Eq.~(\ref{eq:M3Luu}), that with two factors of $\Kdf$. 
The factors of ${\cal L}_L^{(u)}$ and ${\cal R}_L^{(u)}$ on the ends lead to 
the same volume-dependent factors as in Eq.~(\ref{eq:LLKRLdecomp}).
What is new are the finite-volume effects between the two factors of $\Kdf$.
We find
\begin{align}
- \Kdf F_3 \Kdf & = - \Kdf \frac{\rho}{2 \omega L^3} \mathcal L_L^{(u)} \Kdf \\ 
&=  - \Kdf \rho \mathcal M_2 \frac{1}{2 \omega L^3 \mathcal M_2} \mathcal L_L^{(u)} \Kdf  \\ 
& %\hspace{-75pt} 
=  \Kdf \bigg \{\!-\! \mathcal F_3 
\nonumber \\
&\hspace{-40pt}
+ \mathcal R^{(u)}  \mathcal C^{(-1)} \sum_{n=0}^\infty \left[    \mathcal D^{(u,u)} \mathcal C^{(-1)} \right ]^n  \mathcal L^{(u)} \bigg \} \Kdf \,.
\label{eq:KFKfinal}
\end{align}
In the first line we have used the definition of $F_3$, Eq.~(\ref{eq:F3def}),
and in the second we have multiplied and divided by $\M$.
The expression is then ready for our standard manipulation of replacing each sum with an
integral plus a sum-integral difference. After some algebra,
again using the result that terms containing $\mathcal C^{(-1)} (-1/3) \Kdf$ may be dropped,
we find the result (\ref{eq:KFKfinal}).
The new quantity $\mathcal F_3$ is the infinite-volume limit of $F_3$, i.e.
\begin{equation}
 \Kdf \mathcal F_3 \Kdf  \equiv \lim_{\epsilon \to 0} \lim_{L \to \infty}  \Kdf F_3 \Kdf  \,.
\end{equation}

We are now in position to complete the all orders summation. 
To do so we organize the terms order by order in $\mathcal C^{(-1)}$. 
First we note that the sum of all terms with no factors of $\mathcal C^{(-1)}$ gives
\begin{multline}
\mathcal D^{(u,u)} 
+ \mathcal L^{(u)} \sum_{n=0}^\infty \left [- \Kdf   \mathcal F_3  \right ]^n   \Kdf \mathcal R^{(u)}  
\equiv  \mathcal M_3^{(u,u)} \,.
\label{eq:M3uudef}
\end{multline}
Here we have finally given the precise definition of $\mathcal M_{3}^{(u,u)}$.\footnote{%
$\mathcal M_3^{(u,u)}$ may also be defined as the finite $i \epsilon$, $L \to \infty$ limit of 
$\MthL^{(u,u)}$. This infinite-volume object becomes the standard three-to-three scattering 
amplitude upon symmetrization, but since symmetrization is not invertible an independent 
definition is required.}
We then observe that $\mathcal M^{(u,u)}_3$ is also the object that emerges in terms containing
factors of $\mathcal C^{(-1)}$, leading to
\begin{equation}
\mathcal M_{3,L}^{(u,u)} = \sum_{n=0}^\infty \mathcal M_3^{(u,u)} \left[\mathcal C^{(-1)} \mathcal M_3^{(u,u)}  \right ]^n + \cdots \,. 
\end{equation}
Summing the series we find that
$\mathcal M_{3,L}^{(u,u)}$ and $\mathcal M_{3}^{(u,u)} $ 
satisfy the same relation as do $\mathcal D_L^{(u,u)}$ and $\mathcal D^{(u,u)}$ in Eq.~(\ref{eq:DLdecomposed}) above, i.e.
\begin{multline}
\label{eq:MLdecomposed}
\mathcal M_{3,L}^{(u,u)}(\vec p, \vec k)  = \mathcal M_3^{(u,u)}(\vec p, \vec k) \\
 +  \bigg [ \frac{1}{L^3} \sum_{\vec \ell} - \int_{\vec \ell} \bigg ] 
  \mathcal M_3^{(u,u)}(\vec p, \vec \ell)       
  \frac{1}{2 \omega_{\ell} \M( \ell)}     \mathcal M_{3,L}^{(u,u)}(\vec \ell, \vec k) + \cdots \,.
\end{multline}
We stress that this result will hold whenever it is legitimate to treat the $F^{i\epsilon}$ and
``$-1/3$ terms" as subleading.

To use Eq.~(\ref{eq:MLdecomposed}) we follow the same
steps as for $\Kdf=0$,  Eqs.~(\ref{eq:Dpole})-(\ref{eq:almostDE}), except now we
are making no approximations aside from keeping only the leading volume dependence.
Specifically, the pole form (\ref{eq:poleS}) for $\Mth$ implies a similar form for
$\mathcal M_3^{(u,u)}$:
\begin{equation}
\label{eq:poleuu}
\mathcal M^{(u,u)}_3(p,k) \sim - 
\frac{\Gamma^{(u)}( p) \overline \Gamma^{(u)}(  k)}{E^{2} - E_B^{2} } \,.
\end{equation}
The only difference from (\ref{eq:poleS}) is that the residue factors are unsymmetrized.
The unsymmetrized finite-volume amplitude has the corresponding pole form
\begin{equation}
\label{eq:MLpole}
\mathcal M_L^{(u,u)}(\vec k', \vec k) \sim - \frac{ \Gamma_L^{(u)}(k')  
\, \overline \Gamma_L^{(u)} \!(k) }{E^2 - (E_B + \Delta E(L))^2} \,.
\end{equation}
Substituting these in Eq.~(\ref{eq:MLdecomposed}) we find Eqs.~(\ref{eq:residuesmatch}) and
(\ref{eq:almostDE}), except now without the need for the $\Kdf=0$ approximation.
Expanding out (\ref{eq:almostDE}) in powers of $\Delta E(L)$ leads to the desired
result, Eq.~(\ref{eq:DE}).

\section{Determining the bound-state residue factors}
\label{sec:calcGamma}

In this section we study the unsymmetrized, s-wave projected residue factors $\Gamma^{(u)}(k)$ and
$\overline \Gamma^{(u)}(k)$. Expressions for these can be found by deriving a relation to the 
nonrelativistic Schr\"odinger wavefunction. 
The latter is known analytically 
(as reviewed, for example, in Ref.~\cite{BH})
and this leads to an analytic result for the residue factors, 
given in Eq.~(\ref{eq:Gammaresult}) below.

We begin by introducing the three-particle wavefunction
$\psi(\vec r_1, \vec r_2, \vec r_3)$, which satisfies
\begin{multline}
\label{eq:SE}
\bigg[ - \frac{1}{2m} \sum_{i} \frac{\partial^2}{\partial \vec r_i^{\,2}} + \sum_{ij} V(\vec r_i - \vec r_j) \bigg ] \psi(\vec r_1, \vec r_2, \vec r_3) \\
= - \frac{\kappa^2}{m} \psi(\vec r_1, \vec r_2, \vec r_3) \,.
\end{multline}
Here $\vec r_i$ are the coordinates of the individual particles. 
Following MRR we suppose that the particles only interact through pairwise potentials. 
If we restrict attention to the center of mass frame, then one of the coordinates becomes redundant. 
It is convenient to express the wavefunction using Jacobi coordinates
\begin{equation}
\label{eq:JacobiCoord}
\vec x_i = \vec r_j - \vec r_k \,, \  \ 
\vec y_i = \tfrac{1}{\sqrt{3}} (\vec r_j + \vec r_k - 2 \vec r_i )  \,,
\end{equation}
where $ijk$ can be assigned any cyclic permutation of $123$.
$\psi$ can then be expressed as a function of any $\vec x_i, \vec y_i$ pair.

As described by MRR, in the unitary limit, 
Eq.~(\ref{eq:SE}) is approximately solved by the wavefunction
\begin{equation}
\psi(\vec x_3, \vec y_3)  = \sum_{i=1}^3 \phi(R, \alpha_i) \,, 
\label{eq:psitophi}
\end{equation}
where 
\begin{equation}
\phi(R, \alpha) = A \kappa \sqrt{D_0} \frac{K_{i s_0}(\sqrt{2} \kappa R)}{R^2} 
\frac{\sinh(s_0 (\pi/2 - \alpha))}{\sinh(\pi s_0/2) \, \sin(2 \alpha)} \,.
\label{eq:phidef}
\end{equation}
Here the hyperradius $R$ and Delves hyperangles $\alpha_i$ are given by
\begin{align}
R^2 &= \frac{\vec x_i^{\,2} + \vec y_i^{\,2}}{2}   \qquad (i=1,\ 2,\ {\rm or}\ 3) \,,
\label{eq:Rhyper}
\\
\alpha_i &= \tan^{-1}\left(\frac{\vert \vec x_i \vert}{\vert \vec y_i \vert}\right) \,.
\label{eq:alphahyper}
\end{align}
Note that the wavefunction (\ref{eq:psitophi}) depends on fewer variables (four)
than the full complement (six). This is because of the neglect of components with
higher angular momenta than s-wave~\cite{BH}.
The coefficient $D_0$ is 
\begin{equation}
D_0 = -  \frac{4}{27 \cdot 3^{1/4} \pi^{7/2}}   c  \,,
\end{equation}
where $c$ is the constant given earlier in Eq.~(\ref{eq:cdef}), while $A$ is normalization
coefficient discussed following Eq.~(\ref{eq:s0def}).

We have chosen the normalization to be
\begin{equation}
 \frac{1}{6} \int d^3 x_3 d^3 y_3 \; J \; \vert  \psi(\vec x_3, \vec y_3 ) \vert^2 
 = \vert A \vert^2 \,,
\label{eq:psinorm}
\end{equation}
where $J=3\sqrt3/8$ is the Jacobian of the transformation from normal to Jacobi coordinates,
and the $1/6$ is due to our use of identical particles.
This differs from the normalization convention used by MRR:
the wavefunction used here is obtained by multiplying that of MRR by $\sqrt{6/J}$.

The decomposition into three terms in Eq.~(\ref{eq:psitophi}) comes from rewriting
the Schr\"odinger equation in Faddeev form~\cite{BH}.
For example, the Faddeev equation satisfied by the part dependent on $\alpha_3$ is
\begin{multline}
\left( - \frac{\kappa^2}{m} + \frac{1}{m} \frac{\partial^2}{\partial \vec x_3^{\,2}} 
+ \frac{1}{m} \frac{\partial^2}{\partial \vec y_3^{\,2}}   \right) \phi(R, \alpha_3) 
\\ 
= V(\vec x_3)  \psi(\vec x_3, \vec y_3)   \,,
\label{eq:Fadeev3}
\end{multline}
and explicitly involves only the potential between particles 1 and 2.
This is the analog in the Schr\"odinger analysis of considering an unsymmetrized
scattering amplitude, in which the first interaction involves only a specific particle
pair (here 1 and 2). Thus for this part of the wavefunction one can think of particle 3
as the spectator, while the other two parts effect the symmetrization.

We will need the Fourier transform of the wavefunction and its components. 
In terms of the momenta of the individual particles, $\vec p_i$, we use the variables
\begin{equation}
\vec k_{12} = \frac{1}{2} (\vec p_1 - \vec p_2 ) \,, 
\ \ \ \vec k_{3} = \frac{1}{3} \left(\vec p_1 + \vec p_2 - 2 \vec p_3 \right ) \,.
\label{eq:Jacobimomenta}
\end{equation}
Since $\vec P= \sum_i \vec p_i=0$, we can also write $\vec k_3= - \vec p_3$, etc..
The Fourier transform is then
\begin{multline}
\widetilde \psi (\vec k_{12}, \vec k_{3} )  
\equiv  \int d \vec x_3 \int d \vec  y_3 \, J
\\ \times \exp \bigg (\!\!-\! i \sum_i \vec r_i \cdot \vec p_i \bigg) \psi(\vec x_3, \vec y_3) \,,
\label{eq:psiFT}
\end{multline}
with 
\begin{equation}
\sum_i \vec r_i \cdot \vec p_i =  
\vec x_3 \cdot \vec k_{12} - \frac{\sqrt{3}}{2} \vec y_3 \cdot \vec k_{3} \,.
\label{eq:FTphase}
\end{equation}
The normalization of the momentum-space wavefunction is then
\begin{equation}
 \frac{1}{6} \int_{\vec k_{12}} \int_{\vec k_{3}} \vert \widetilde \psi(\vec k_{12}, \vec k_{3} ) \vert^2 = \vert A \vert^2 \,.
\end{equation}
A similar definition is used for $\phi$, e.g. 
\begin{multline}
\widetilde \phi_3 (\vec k_{12}, \vec k_{3} )  
\equiv  \int d \vec x_3 \int d \vec  y_3 \, J
\\ \times \exp \bigg (\!\!-\! i \sum_i \vec r_i \cdot \vec p_i \bigg) \phi(R,\alpha_3) \,,
\end{multline}
for the component in which particle 3 is the spectator.

As shown in Appendix \ref{app:details}, 
the residue factor $ \Gamma^{(u)}(\vec k)$ is related to the 
one component of the wavefunction 
(with our normalization) via
\begin{equation}
\label{eq:Gammafrompsi}
 \frac{\Gamma^{(u)}( k)}{4\sqrt3 m^2}
=
 \lim_{\mathrm{on\ shell}}
 \left (-\frac{\kappa^2}{m}-H_0 \right ) \widetilde \phi_3(\vec k_{12}, \vec k_{3}) \,,
\end{equation}
where
\begin{equation}
H_0 = \sum_{i=1}^3 \frac{\vec p_i^2}{2 m} \,.
\label{eq:H0def}
\end{equation}
The on-shell limit is effected by setting $\vec k_{3} = -\vec p_3 \to - \vec k$ 
(the spectator momentum) and sending
 $\vec k_{12}$ to a (complex) value such that $H_0=-\kappa^2/m$.
As we will see by explicit calculation, the result for $\Gamma^{(u)}$ only depends on $k=|\vec k|$, and
not on the direction of $\vec k$ nor on the remaining on-shell angular variable $\hat a^*$
(defined in Sec.~\ref{sec:setup}). This is expected since the s-wave projection removes dependence on $\hat a^*$. Dependence on $\hat k$ is also removed as this can only appear in a scalar product, and no other directions are defined for $\Gamma^{(u)}$ in the CM frame.

To evaluate the right-hand side of Eq.~(\ref{eq:Gammafrompsi}), we first return to
position space. As we show in Appendix \ref{app:identity}, given the explicit form for $\phi$,
Eq.~(\ref{eq:phidef}), one can derive the identity
\begin{multline}
\left( - \frac{\kappa^2}{m} + \frac{1}{m} \frac{\partial^2}{\partial \vec x_3^{\,2}} 
+ \frac{1}{m} \frac{\partial^2}{\partial \vec y_3^{\,2}}   \right) \phi(R, \alpha_3) 
\\ 
= - 4 \pi A  \sqrt{D_0} \frac{\kappa}{m}  
\frac{K_{i s_0}(\kappa \vert \vec y_3 \vert)}{\vert \vec y_3 \vert} \delta^3(\vec x_3) \,.
\label{eq:identity}
\end{multline}
Comparing to the Faddeev equation (\ref{eq:Fadeev3}) we see that the approximate form
of the wavefunction we are using corresponds to a potential proportional to a delta function.
Fourier transforming, we use Eq.~(\ref{eq:Gammafrompsi}) to obtain
\begin{multline}
\label{eq:GammauFT}
 \Gamma^{(u)}( k) =  -4 \sqrt{3}   m^2   \int  \! d^3x_3   
  \! \int   \! d^3y_3 \,J\, \exp \bigg (\!\!-\! i \sum_i \vec r_i \cdot \vec p_i \bigg) 
  \\ 
  \times 4 \pi A  \sqrt{D_0} \frac{\kappa}{m}  \frac{K_{i s_0}
  (\kappa \vert \vec y_3 \vert)}{\vert \vec y_3 \vert} \delta^3(\vec x_3) \,.
\end{multline}

We have not specified the on-shell limit because it turns out to be trivial.
Using the form of the Fourier transform phase given in Eq.~(\ref{eq:FTphase}),
and setting $\vec k_3=-\vec k$, we see that the trivial $\vec x_3$ integral
removes all dependence on $\vec k_{12}$:
\begin{multline}
\Gamma^{(u)}(k) = -4 \sqrt{3} m^2 4 \pi  A \sqrt{D_0}      \frac{\kappa}{m}  J \\ \times
  \int d^3y_3   \frac{K_{i s_0}(\kappa \vert \vec  y_3 \vert)}{ \vert \vec  y_3 \vert} e^{i \sqrt{3} \vec  y_3 \cdot \vec k /2 } \,.
\end{multline}

Substituting the value of $\sqrt{D_0}$ and simplifying then gives
\begin{equation}
\Gamma^{(u)}(k) = -\frac{4 \cdot 3^{3/8} }{\pi^{3/4}} A \sqrt{-c} \, \kappa m     \,  g(\sqrt{3} k/2) \,,
\end{equation}
where
\begin{equation}
g(q) \equiv \int \! 2\pi d(\cos\theta) y_3^2 d y_3  \frac{K_{i s_0}(\kappa   y_3 )}
{  y_3 } e^{i  y_3  q \cos \theta } \,.
\end{equation}
Evaluating the angular integral then gives
\begin{equation}
g(q) =  \frac{4\pi}{\kappa q} \int_0^\infty dz \sin(z q/\kappa)\, K_{i s_0}(z) \,.
\end{equation}
The remaining integral may also be evaluated analytically:
\begin{multline}
g(q) = \frac{2\pi^2}{\kappa^2 \sinh(s_0\pi/2)}  \\ \times \frac{\sin[s_0\sinh^{-1}(q/\kappa)]}{q/\kappa}
\frac1{\sqrt{1+q^2/\kappa^2}}
\,.
\end{multline}
The function $g(q)$ is singular at $q^2=-\kappa^2$, due to both factors on the second line. 
The expansion about the singular point has the form
\begin{multline}
g(q) =  \frac{2\pi^2}{\kappa^2} \bigg [ \frac1{\sqrt{1+q^2/\kappa^2}}
- s_0 \coth(s_0\pi/2) \\ + {\cal O}\Big(\sqrt{1+q^2/\kappa^2}\Big) \bigg ]
\,.
\end{multline}
The leading singularity will lead to the dominant finite-volume effects.

We conclude that the leading contribution to $\Delta E(L)$ is given by evaluating the expression derived in the previous section [Eq.~(\ref{eq:DE})] with
\begin{equation}
\label{eq:Gammaresult}
\Gamma^{(u)}(k) = -8 \cdot 3^{3/8} \pi^{5/4}       A   \sqrt{-c}      \frac{m}{\kappa}  \left[1 + \frac{3 k^2}{4 \kappa^2} \right ]^{-1/2} \,.
\end{equation}
Repeating the exercise for $\overline \Gamma^{(u)}(k)$, one finds
the same form up to a complex conjugate which has no effect other than $A \to A^*$. 

\section{Determination of $\Delta E(L)$}
\label{sec:calcDE}

In this section we use the result for $\Gamma^{(u)}$ derived in the previous section
to evaluate the energy shift $\Delta E(L)$ using Eq.~(\ref{eq:DE}).

To do so we need the expression for $\M^s(k)$ in the unitary limit.
This requires the kinematic quantities $E_{2,k}^*$ and $q_k^*$ 
[defined in Eqs.~(\ref{eq:E2kdef}) and (\ref{eq:qkdef}), respectively]
\begin{align}
\begin{split}
E_{2,k}^* &= \sqrt{ (3m - \kappa^2/m - \omega_k)^2 - k^2}\,, \\
&= 2m \left[1 - (\kappa^2 + 3 k^2/4)/(2m^2) + \cdots\right]\,,\\
q_k^* &= \sqrt{- \kappa^2 - 3 k^2/4 + \cdots} \,,
\end{split}
\label{eq:kinematics}
\end{align}
with the ellipses indicating higher-order terms in the nonrelativistic expansion
(assuming $k^2\sim \kappa^2 \ll m^2$).
We see that the momentum $q_k^*$ is pure imaginary, 
as expected since we are studying a subthreshold energy. 
Below threshold the scattering amplitude takes the form
\begin{equation}
\label{eq:M2result0}
\mathcal M_2(k) = \frac{16 \pi E_{2,k}^*}{q_k^* \cot \delta(q_k^*) + \vert q_k^* \vert} \,,
\end{equation}
where we use
$q_k^*\cot \delta(q_k^*) = -1/a  + r (q_k^*)^2/2 + \cdots$,
with $a$ the scattering length and $r$ the effective range,
to perform the analytical continuation.
In the unitary limit, $a\to -\infty$, we therefore have
\begin{equation}
\label{eq:M2result1}
\mathcal M_2(k) \longrightarrow 
\frac{16 \pi E_{2,k}^*}{\vert q_k^* \vert}\left[1 + {\cal O}(q_k^* r)  \right]\,.
\end{equation}
Inserting the results from Eq.~(\ref{eq:kinematics}), and dropping suppressed 
terms, gives the form we need for our computation:
\begin{equation}
\label{eq:1byM2}
\frac{1}{\mathcal M_2(k)} = \frac{\kappa}{32 \pi m} 
\left[1 + \frac{3 k^2}{4 \kappa^2}   \right ]^{1/2} + \cdots \,.
\end{equation}

The energy shift, Eq.~(\ref{eq:DE}), can now be written, using
Eqs.~(\ref{eq:Gammaresult}) and (\ref{eq:1byM2}), as
\begin{multline}
\Delta E(L) = c \vert A \vert^2 64 \cdot 3^{3/4} \pi^{5/2} \frac{m^2}{\kappa^2} \, \frac{\kappa}{32 \pi m} \, \frac{1}{2 E_B} \\ \times \bigg[ \frac1{L^3} \sum_{\vec k} - \int_{\vec k} \bigg] \frac{1}{2 \omega_k}   \left[1 + \frac{3 k^2}{4 \kappa^2}   \right ]^{-1/2} + \cdots \,.
\label{eq:DE1}
\end{multline}
Applying the Poisson summation formula we find
\begin{multline}
\label{eq:DEpoisson}
\Delta E(L) = c \vert A \vert^2 \frac{\pi^{3/2}}{3^{1/4}} \frac{1}{\kappa} 
\sum_{\vec s\ne 0}  
\\ \times 
\int \frac{d^3 k}{(2 \pi)^3} e^{i L \vec s \cdot \vec k} \frac{1}{2 \omega_k}   
\left[1 + \frac{3 k^2}{4 \kappa^2}   \right ]^{-1/2} +\cdots \,,
\end{multline}
with $\vec s$ a vector of integers.
We have also set $E_B = 3m$, which holds up to corrections down by $\kappa^2/m^2$ . 
We further simplify by evaluating the angular integral 
and using the symmetry of the resulting integrand to 
extend the $k$ integral to the entire real axis:
\begin{multline}
\label{eq:beforerotation}
\Delta E(L) = c \vert A \vert^2 \frac{3^{3/4}}{24 \sqrt{\pi}  i \kappa m L} 
\sum_{s=1}^\infty \frac{\nu_s}s
\\ \times \int_{- \infty}^\infty \!\! dk   k \,  e^{i s k L}  
 \left [ \left (1 + \frac{k^2}{m^2} \right ) \left (1 + \frac{3 k^2}{4 \kappa^2} \right ) \right ]^{-1/2} 
 +\cdots\,,
\end{multline}
where $\nu_s$ is the number of integer vectors $\vec s$ with magnitude $s$ 
(e.g.~$\nu_1=6,\ \nu_2=12,\ \cdots$). 
We next deform the contour so as to wrap around the branch cut 
along the positive imaginary axis, and introduce $\ell = \pm i k$ to 
parametrize the integral along the discontinuity,
\begin{multline}
\label{eq:introtated}
\Delta E(L) = c \vert A \vert^2 \frac{3^{3/4}}{12 \sqrt{\pi}   \kappa m L}\sum_{ s=1}^\infty \frac{\nu_s}s
 \\ 
 \times  \int_{2 \kappa/\sqrt{3}}^\infty d\ell  \ell  \,  e^{- s \ell L}    
 \left ( \frac{3 \ell^2}{4 \kappa^2} - 1 \right )^{-1/2} + \cdots\,.
\end{multline}
Note that at this stage we have set $(1 + k^2/m^2)^{-1/2}$ to unity. 
This factor is required to ensure convergence of the integral but, 
after the contour has been deformed, 
it can be expanded in powers of $k^2/m^2$. 
Upon integration these contribute subleading powers of $\kappa^2/m^2$ that we neglect.

It is now apparent that the integral will have an exponential fall-off proportional to
$\exp(-s2\kappa L/\sqrt3)$. Thus we need only keep the sixfold degenerate 
$s=1$ term. Doing so, and evaluating the integral, we reach
\begin{equation}
\Delta E(L) =  c \vert A \vert^2  \frac{\kappa^2}{m} \frac{2}{3^{1/4} \sqrt{\pi} \kappa L}  \, K_1 \! \left(\frac{2 \kappa L}{\sqrt{3}}\right) + \cdots \,.
\end{equation}
Substituting the asymptotic form of the Bessel function we obtain the MRR result,
Eq.~(\ref{eq:result}).

\section{Comparison with two-particle bound-state energy shift}
\label{sec:twopart}

In this section we compare the result just obtained,
along with its derivation, with the corresponding result and derivation for
the energy shift for a spin-zero two-particle bound state, $\Delta E_2(L)$.

The leading-order volume dependence of $\Delta E_2(L)$
has been quoted in Eq.~(\ref{eq:result2}). We first recall the standard
derivation of this result. This uses
L\"uscher's quantization condition (assuming s-wave dominance),
which in our notation reads~\cite{KSS}
\begin{equation}
1/\M^s + F_2^{i\epsilon}(E_2,\vec 0) = 0\,.
\label{eq:KSS}
\end{equation}
The two-particle zeta-function for total momentum $\vec P$ is given by
\begin{multline}
\label{eq:F2ieps}
F_2^{i\epsilon}(E_2,\vec P)
 \equiv \frac12 \bigg[\frac{1}{L^3} \sum_{\vec k} - \int_{\vec k} \bigg]  
 \\ \times
\frac{   H(\vec k) H(\vec P-\vec k)}
{2 \omega_k 2 \omega_{kP}(E - \omega_k  -  \omega_{k P} + i \epsilon)}
  \,,
\end{multline}
where $\omega_{kP} = \sqrt{m^2 + (\vec P-\vec k)^2}$.
$F_2^{i\epsilon}$ is related to the function $F^{i\epsilon}(\vec \ell)$ defined in 
Eq.~(\ref{eq:Fiepsaltdef}) by
\begin{equation}
F_2^{i\epsilon}(E_2,\vec P) = F^{i\epsilon}(-\vec P)\bigg|_{E-\omega_P=E_2}
\,.
\end{equation}
Here we consider a state at rest, and so set $\vec P=0$.
We parametrize the two-particle energy as $E_2=2m -\kappa_2^2/m$,
with $\kappa_2$ at this stage arbitrary except that $\kappa_2\ll m$.
Then we have~\cite{Beane:2003da}\footnote{%
This result agrees with that from Eq.~(\ref{eq:FiepsNR}).}
\begin{equation}
F_2^{i \epsilon}(E_2,\vec 0)  = - \frac{3}{16 \pi m L } e^{- \kappa_2 L} \,,
\label{eq:Fiepsrest}
\end{equation}
up to terms suppressed by $\kappa_2^2/m^2$.
The scattering amplitude is given by Eq.~(\ref{eq:M2result0}), 
where now $|q_k^*|=\kappa_2$ and $E_{2,k}^*\approx 2m$. 
Using the effective range expansion, and assuming that the scattering length
dominates,\footnote{%
Since here we are studying momenta $q_k^* \sim \kappa_2 = 1/a$, the effective
range term in $q_k^*\cot\delta$ is suppressed by a relative factor of $r/a$.}
one finds
\begin{equation}
\frac1{\M^s} = \frac{\kappa_2 - 1/a}{32\pi m}\,.
\label{eq:1byM2s}
\end{equation}
$\M^s$ thus has a pole when $\kappa_2=1/a$, 
corresponding to the bound-state energy in infinite volume
of $E_{B_2}=2m - 1/(a^2m)$.
Inserting (\ref{eq:1byM2s}) into the quantization condition (\ref{eq:KSS}),
and using the fact that the energy shift is small, we find
\begin{equation}
\Delta\kappa_2(L) =  \frac{6}{L} e^{-L/a}
\,,
\end{equation}
where
\begin{equation}
E_{B_2}(L) = E_{B_2} + \Delta E_2(L) = 
2 m - \frac{[\kappa_2 + \Delta \kappa_2(L)]^2}{m}\,.
\end{equation}
This leads to the result quoted earlier, Eq.~(\ref{eq:result2}).

We now repackage this derivation using a method analogous to that used for three particles.
We consider the two-particle finite-volume scattering amplitude, $\ML$,
which satisfies
\begin{equation}
\ML = \M^s - \M^s F^{i \epsilon} (\vec 0)\ML \,,
\end{equation}
as can be seen from Eq.~(\ref{eq:MLdef}). 
This is the analog of Eq.~(\ref{eq:MLdecomposed}), except that here there are no
subleading sources of finite-volume dependence (except those proportional to $e^{-mL}$,
which are dropped throughout).
Substituting the pole ans\"atze
\begin{align}
\label{eq:M2pole}
 \M & = - \frac{ \Gamma_2 \overline  \Gamma_2}{E^{2} - E_{B,2}^2} \,, \\
 \ML & = - \frac{ \Gamma_{2,L}  \overline \Gamma_{2,L}}{E^{2} - [E_{B_2}+ \Delta E_2(L)]^2} \,, 
\end{align}
and following steps analogous to Eqs.~(\ref{eq:Dpole})-(\ref{eq:almostDE}) above, we find
\begin{equation}
\label{eq:twobound}
  \Delta E_2(L) =   \frac{1}{2 E_{B_2}} 
\overline \Gamma_2 F_2^{i \epsilon}(E_{B_2},\vec 0)\Gamma_2\,.
\end{equation}
This is the analog of Eq.~(\ref{eq:DE}), except that here the residues are numbers rather
than functions and the sum-integral difference is not explicit but instead included in the 
definition of $F_2^{i \epsilon}$.\footnote{%
We can cast the result into a form even more similar to Eq.~(\ref{eq:DE})
by using the quantization condition to write $F^{i\epsilon}=-1/\M^s$.
This result holds when $E_2=E_{B_2}+\Delta E_2(L)$, which is an equally 
valid choice for the energy at which
to evaluate the right-hand side of Eq.~(\ref{eq:twobound}).
However, this substitution leads only to the vacuous result
$\Delta E_{2, \vec P}(L) = \Delta E_{2, \vec P}(L)$ and is thus not useful.}
The residues can be obtained by matching the pole ansatz for $\M$, Eq.~(\ref{eq:M2pole}),
with the specific result (\ref{eq:1byM2s}), leading to\footnote{%
It is also possible to derive this result from the Schr\"odinger wavefunction
for a two-particle weakly bound state using an analog of the relation
Eq.~(\ref{eq:Gammafrompsi}).}
\begin{equation}
\overline \Gamma_2  \Gamma_2  = 256 \pi m/a \,.
\label{eq:residues2}
\end{equation}
Inserting this into the new form for $\Delta E_2(L)$,
Eq.~(\ref{eq:twobound}), along with the result (\ref{eq:Fiepsrest}) for $F_2^{i\epsilon}$,
we find again the energy shift quoted in Eq.~(\ref{eq:result2}).

We now compare the recast two-particle result (\ref{eq:twobound}) with the three-particle
result, Eq.~(\ref{eq:DE}), in more detail.
Both have a form analogous to a leading-order correction in
perturbation theory: a ``matrix element'' evaluated between unperturbed ``wavefunctions''.
The ``operator'' in both cases involves a sum-integral difference---this is explicit
in Eq.~(\ref{eq:DE}) and contained in $F_2^{i\epsilon}$ in the two-particle case.
This is expected, since it is the difference between
sums and integrals that leads to finite-volume effects.
The results differ in the nature of the process occurring in the sum-integral difference.
For two particles, it is just a subthreshold loop of two free particles, as can be seen from
the form of $F_2^{i\epsilon}$, or by returning to the original derivation, e.g. in Ref.~\cite{KSS}.
For three particles this does not simply generalize to a subthreshold three-particle loop---such
loops give rise to the $F^{i\epsilon}$ terms that are shown in Appendix \ref{app:approx} to
be subleading by a factor of $1/L$.
Instead, what appears is a process in which two of the three
particles are scattering. This leads to
the appearance of the explicit factor of $1/\M^s$ in Eq.~(\ref{eq:DE}),
as well as to the singularities in the residues.

\section{Generalization to a moving bound state}
\label{sec:moving}

In this section we extend the result derived above to the case where the
three-particle bound state 
has nonzero momentum, $\vec P$, in the finite-volume frame. 
This momentum is constrained by the boundary conditions to satisfy 
$\vec P=(2\pi/L)\vec n_P $,
with $\vec n_P$ a vector of integers.
We study the case in which $\vec n_P$ is fixed, so that $\vec P \sim 1/L \ll m$.
The alternative in which one holds $\vec P \sim m$ is also interesting
(since it more closely approximates moving frames used in present simulations)
but this leads to more complicated expressions and goes beyond the scope of this work. 

Generalizing to nonzero momentum turns out to be straightforward. 
We define the energy shift to be that in the CM-frame bound-state energy,
so that the energy of the FV state in the moving frame is
\begin{equation}
E_B(\vec P, L) \equiv \sqrt{[E_B + \Delta E_{\vec P}(L)]^2 + \vec P^2} \,.
\end{equation}
The steps of Sec.~\ref{sec:QCexp} go through unchanged,\footnote{%
The only caveat is that the arguments given in Appendix \ref{app:approx} 
that certain terms are suppressed at large $L$ need to be reconsidered.
They continue to hold when $\vec P \sim 1/L$, the case considered here,
but it is unclear whether they hold when $\vec P\sim m$.}
and one arrives again at Eq.~(\ref{eq:MLdecomposed}).
The only subtlety is that the energy of the nonspectator pair now depends
also on $\vec P$. If the spectator momentum is $\vec k$, then
the two-particle CM frame energy becomes
\begin{equation}
E_{2,k}^*(\vec P) = \sqrt{(E - \omega_{k})^2 - (\vec P - \vec k)^2} \,,
\end{equation}
and the individual CM frame momenta are
\begin{equation}
q_k^*(\vec P) = \sqrt{ E_{2,k}^{*}(\vec P)^2/4 - m^2}\,.
\end{equation} 
Since the quantities that enter the finite-volume energy shift, i.e.~$\M^s$, $\Gamma^{(u)}$ and $\overline\Gamma^{(u)}$, are Lorentz scalars, expressing these as functions of $q_{k}^*$, rather than of $\vec k$, results in expressions that hold in all frames.\footnote{%
The correspondence to the previous notation is, for example,
$\Gamma^{(u)}[q_k^*(\vec 0)] = \Gamma^{(u)}(k)$.}
For example, the generalization of the pole form for $\Mth^{(u,u)}$, 
Eq.~(\ref{eq:poleS}), is
\begin{equation}
\label{eq:poleP}
\mathcal M_3(\vec p; \vec k) 
\sim - \frac{\Gamma[q_{p}^*(\vec P)] 
\overline \Gamma[q_{k}^*(\vec P)]}{E^{*2} - E_B^{2}} \,,
\end{equation}
where $E^{*2}=E^2 - \vec P^2$.
A similar form holds for $\MthL^{(u,u)}$, except with $E_B$ replaced with
$E_B+ \Delta E_{\vec P}(L)$.

Substituting these pole forms into Eq.~(\ref{eq:MLdecomposed}), and proceeding as before,
we find
\begin{multline}
\label{eq:movingDE}
\Delta E_{\vec P}(L) = - \frac{1}{2 E_B} \bigg[ \frac1{L^3} \sum_{\vec k} - \int_{\vec k} \bigg] 
\frac1{2\omega_k}
\\ 
\times  \frac{\overline \Gamma^{(u)} \![q_{k}^*(\vec P)] \   
\Gamma^{(u)}[q_{k}^*(\vec P)] }{\mathcal M_{2}[q_{k}^*(\vec P)]} + \cdots \,.
\end{multline}
This is to be evaluated at the
infinite-volume moving-frame bound-state energy,
\begin{equation}
E_B(\vec P)=\sqrt{E_B^2+ \vec P^2} \,.
\end{equation}
We note that $\M^s$ is already expressed in terms of Lorentz scalars in Eq.~(\ref{eq:M2result1}),
while $\Gamma^{(u)}$ in Eq.~(\ref{eq:Gammaresult}) can be rewritten in invariant form
using Eq.~(\ref{eq:kinematics}):
\begin{equation}
\label{eq:Gammaresult2}
\Gamma^{(u)}[q_k^*(\vec P)] = - 8 \cdot 3^{3/8} \pi^{5/4}       A   \sqrt{-c}      
\frac{m}{|q_k^*(\vec P)|} \,.
\end{equation}
Since $\int_{\vec k}(1/\omega_k)$ is also invariant, we see that the only noninvariant part of
the expression for $\Delta E_{\vec P}(L)$ is the sum.

Applying the Poisson summation formula, and dropping terms suppressed by powers of
$\kappa$, we find
\begin{multline}
\label{eq:DEpoissonmove}
\Delta E_{\vec P}(L) = c \vert A \vert^2 \frac{\pi^{3/2}}{3^{1/4}}  \sum_{\vec s}    
 \int \frac{d^3 k}{(2 \pi)^3} \frac{e^{i L \vec s \cdot \vec k}}{2 \omega_{k}}   
\frac1{|q_k^*(\vec P)|} + \cdots \,.
\end{multline}
To further simplify we change the variable of integration to $\vec k^*$, 
defined by boosting the four-vector $(\omega_k, \vec k)$ to the three-particle CM frame. 
The only non-invariant factor is the exponent, and this can be written
\begin{align}
\vec s \cdot \vec k & = \vec s \cdot \left(\vec k^* + \frac{\omega_{k^*}}{E_B}\vec P \right) 
+
\left(\gamma - 1 \right ) \frac{(\vec s \cdot \vec P)( \vec k^*  \cdot \vec P)}{\vec P^2} \,,
\end{align}
where $\gamma=E_B(\vec P)/E_B$. Since we are scaling $\vec P$ as $1/L$, we can set 
$\gamma=1$ and drop the last term. Also, since the integral is dominated by
nonrelativistic momenta, and given that $\kappa \ll m$, 
we can set $\omega_{k^*}/E_B =1/3$. Thus we arrive at
\begin{equation}
\vec s \cdot \vec k = \vec s \cdot \left( \vec k^* + \frac{\vec P}{3}\right)
\left[1 + \mathcal O\left(\frac1{(mL)^2}, \frac{\kappa^2}{m^2}\right) \right]\,.
\end{equation}
 This is the result that one expects from a Galilean boost,
in which each of the three particles picks up momentum $\vec P/3$. 
Substituting this into Eq.~(\ref{eq:DEpoissonmove}), we reach 
\begin{multline}
\label{eq:DEchangeframe}
\Delta E_{\vec P}(L) = c \vert A \vert^2 \frac{\pi^{3/2}}{3^{1/4}} \frac1{\kappa} \sum_{\vec s}   
e^{i(2\pi/3) \vec s \cdot \vec n_P} 
\\ \times
 \int \frac{d^3 k^*}{(2 \pi)^3} \frac{e^{i L \vec s \cdot \vec k^*}}{2 \omega_{k^*}}   
 \left[ 1 + \frac{3 k^{*2}}{4\kappa^2}   \right ]^{-1/2} + \cdots \,.
\end{multline}

We now observe that the integral appearing in (\ref{eq:DEchangeframe}) is
{\em identical} to that in the rest-frame expression (\ref{eq:DEpoisson}).
Indeed, the only difference between the expressions is the presence of
the phase factor $\exp[{i(2\pi/3) \vec s\cdot \vec n_P}]$ in (\ref{eq:DEchangeframe}).
Keeping only the dominant $s=1$ terms in the Poisson sum we 
find that the energy shifts in different frames are related by a simple prefactor 
\begin{equation}
\Delta E_{\vec P}(L) = f_3[\vec n_P]\Delta E(L) + \cdots \,,
\label{eq:DEP}
\end{equation}
with
\begin{equation}
f_3[{\vec n_P}] = \frac{1}{6} \sum_{\hat s }  e^{i (2\pi/3) \hat s \cdot \vec n_P }\,.
\label{eq:prefactor3}
\end{equation}
We stress that the sum here is only over the six unit vectors $\hat s$.
This prefactor varies dramatically with the value of momentum. 
For example, the lowest momenta give
\begin{equation}
\begin{split}
f_3[{(0,0,0)}]  &= 1 \,, \ \  \ \ 
f_3[{(0,0,1)}]  = 1/2\,, 
\\ 
f_3[{(0,1,1)}]  &= 0 \,,  \ \ \ \ 
f_3[{(1,1,1)}]  = -1/2 \,.
\end{split}
\label{eq:prefactorexs}
\end{equation}

This result is very similar to that for a two-particle bound state,
as described in Ref.~\cite{Davoudi:2011md}.
We can obtain the results in our approach by noting that
Eq.~(\ref{eq:twobound}) generalizes to
\begin{equation}
\label{eq:QCtwo}
  \Delta E_{2, \vec P}(L) =   
  \frac{1}{2 E_{B_2}} \overline \Gamma_2 F_2^{i \epsilon}(E_{B_2}(\vec P),\vec P) 
  \Gamma_2 \,,
\end{equation}
where $E_{B_2}(\vec P)^2=E_{B_2}^2+\vec P^2$.
The residues are independent of $\vec P$ and given by Eq.~(\ref{eq:residues2}).

Assuming fixed $\vec n_P$ as $L\to\infty$,
the leading-order form for $F_2^{i \epsilon}(\vec P)$ is~\cite{Davoudi:2011md}
\begin{equation}
\label{eq:Fiepsresult}
F_2^{i \epsilon}(E_{B_2}(\vec P),\vec P)  
= - \frac{1}{32 \pi m L } e^{- \kappa_2 L}  \sum_{\hat s }  e^{i \pi \vec n_P \cdot \hat s  } + \cdots  \,,
\end{equation}
where again the sum over $\hat s$ runs over the six unit vectors.
It follows that the energy shifts for different $\vec n_P$ 
have a similar form to the three-particle case:
\begin{align}
\Delta E_{2,\vec P}(L) 
& = f_2[{\vec n_P}] \Delta E_2(L) + \cdots \,,
\end{align}
where 
\begin{align}
f_2[\vec n_P] &= \frac16 \sum_{\hat s} e^{i\pi \hat s \cdot \vec n_P}\,.
\end{align}
The values of $f_2$ for the lowest momenta are
\begin{equation}
\begin{split}
f_2[{(0,0,0)}]  &= 1 \,, \ \  \ \  \ \ \ \ \
f_2[{(0,0,1)}]  = 1/3\,, 
\\ 
f_2[{(0,1,1)}]  &= -1/3 \,,  \ \ \ \ 
f_2[{(1,1,1)}]  = -1 \,.
\end{split}
\end{equation}

\section{Conclusions}
\label{sec:conc}

The main motivation for this work was to provide a further nontrivial
check of our three-particle quantization condition. 
While many technical steps are required to carry out this check, 
the key result for the energy shift, Eq.~(\ref{eq:DE}), is rather simple.
We have derived this result for a particular type of three-particle bound state,
namely a spin-zero state for which the two-particle interaction is near the unitary limit.
It would be interesting to know, however, whether Eq.~(\ref{eq:DE}) 
gives the leading volume dependence in a more general context, 
or whether contributions that are higher order here, such as $\Delta E_F(L)$
in Eq.~(\ref{eq:DEL}), must be considered.

Our extension of the result for the energy shift to a moving frame shows the
utility of having a formalism that holds for any momentum $\vec P$. 
It also opens up the possibility of generalizing the work of  Ref.~\cite{Davoudi:2011md} 
from two- to three-particle bound sates. The idea is to
determine linear combinations of three-body bound-state energies 
(obtained from different frames) for which the leading finite-volume dependence cancels. 

Indeed, from our results here it is already clear that such a cancellation occurs
if one averages the CM-frame energies extracted from the $\vec P = (2 \pi/L)(0,0,1)$
and $\vec P = (2 \pi/L)(1,1,1)$ frames. 
Even more striking is the observation that the leading finite-volume effects 
vanish for the $\vec P = (2 \pi/L)(0,1,1)$ frame, implying that energies 
extracted in this frame are closer to the infinite-volume 
three-particle bound-state energy than those obtained in the rest frame.
It is important to keep in mind, however, that here the
subleading terms that are not canceled are suppressed only by a power of $1/(\kappa L)$,
whereas those in the two-particle case are exponentially suppressed.

%%%%%%%%%%%%%%%%%%%%%%%%%%%%%%%%%%%%%%%%%%%%%
%%%%%%%%%%%%%%%%%%%%%%%%%%%%%%%%%%%%%%%%%%%%%
%%%%%%%%%%%%%%%%%%%%%%%%%%%%%%%%%%%%%%%%%%%%%
%%%                                                                                                                             %%%
%%%                                            THE APPENDICES                                                  %%%
%%%                                                                                                                             %%%
%%%%%%%%%%%%%%%%%%%%%%%%%%%%%%%%%%%%%%%%%%%%%
%%%%%%%%%%%%%%%%%%%%%%%%%%%%%%%%%%%%%%%%%%%%%
%%%%%%%%%%%%%%%%%%%%%%%%%%%%%%%%%%%%%%%%%%%%%
\section*{Acknowledgments}
We thank Ulf Mei{\ss}ner and Akaki Rusetsky for discussions and correspondence.
The work of SS was supported in part by the United States Department of Energy 
grant DE-SC0011637. SS thanks the Institut f\"ur Kernphysik
 and Helmholtz Institut Mainz
for hospitality while some of this work was completed.

%\newpage
\appendix

\section{Justifying approximations}

\label{app:approx}

In this appendix we justify various approximations used in the main text.
We do so in three steps. First, we show that the FV effects in $\Delta E(L)$ arising from 
factors of $F^{i\epsilon}$ are subleading in the calculation for $\Kdf=0$. 
Second, we argue that the same holds for the calculation of
$\Delta E(L)$ with nonzero $\Kdf$. Finally, we argue that the $-1/3$ terms
contained in Eq.~(\ref{eq:LKR}), and implicitly
in Eq.~(\ref{eq:KFKfinal}),  also lead to subleading corrections to $\Delta E(L)$.

\subsection{Dropping $F^{i\epsilon}$ terms if $\Kdf=0$}

We start from the general form of $\mathcal D_L^{(u,u)}$ [Eq.~(\ref{eq:DLdef})]
and use the result (\ref{eq:MLdef}) to expand $\ML$ in powers of $F^{i\epsilon}$.
After some algebra
we find the following matrix equation 
\begin{multline}
\left([2\omega L^3] \ML + \mathcal D_L^{(u,u)}\right)
=
\left([2\omega L^3] \M^s + \mathcal D_G^{(u,u)}\right)
\\
- 
\left([2\omega L^3] \M^s + \mathcal D_G^{(u,u)}\right) \frac{F^{i\epsilon}}{[2\omega L^3]}
\left([2\omega L^3] \ML + \mathcal D_L^{(u,u)}\right)
\,,
\label{eq:DLvsDG}
\end{multline}
where
\begin{equation}
\label{eq:DGdef}
\mathcal D^{(u,u)}_G \equiv -\frac{1}{1 + \M G} \M G [2 \omega L^3] \M \,,
\end{equation}
is simply the approximation for $\mathcal D^{(u,u)}_L$ used
in Sec.~\ref{subsec:Kzero}, i.e.~Eq.~(\ref{eq:DLred1}).
It satisfies Eq.~(\ref{eq:DLdecomposed}) without approximation:
\begin{multline}
\label{eq:DGdecomposed}
\mathcal D_G^{(u,u)}(\vec k, \vec p)  = \mathcal D^{(u,u)}(\vec k, \vec p) \\
 +  \bigg [ \frac{1}{L^3} \sum_{\vec \ell} - \int_{\vec \ell} \bigg ]  
\mathcal D^{(u,u)}(\vec k, \vec \ell)  
     \frac{1}{2 \omega_{\ell} \M( \ell)}     \mathcal D_G^{(u,u)}(\vec \ell, \vec p)\,.
\end{multline}

We now repeat in two stages the argumentation given at the end of Sec.~\ref{subsec:Kzero}.
First we use the pole form for $\mathcal D^{(u,u)}$, Eq.~(\ref{eq:Dpole}), 
which,  using Eq.~(\ref{eq:DGdecomposed}), implies that $\mathcal D_G^{(u,u)}$ has the
pole form 
\begin{equation}
\label{eq:DGpole}
\mathcal D_G^{(u,u)}(\vec k', \vec k)\bigg|_{\Kdf=0} \sim - \frac{ \Gamma^{(u)}(k')  
\, \overline \Gamma^{(u)} \!(k) }{E^2 - (E_B + \Delta E_G(L))^2} \,,
\end{equation}
with the energy shift
\begin{equation}
\label{eq:DEG}
\Delta E_G(L) = - \frac{1}{2 E_B} 
\bigg[ \frac1{L^3} \sum_{\vec k} - \int_{\vec k} \bigg]  
\frac{\overline \Gamma^{(u)} \!(k) \,  \Gamma^{(u)}(k)  }{2 \omega_k \mathcal M_2^s(k)}
+ \cdots \,.
\end{equation}
This is the energy shift (\ref{eq:DE}) determined in Sec.~\ref{subsec:Kzero}
in the approximation of dropping factors of $F^{i\epsilon}$.

The second stage is to substitute the pole forms for
$\mathcal D_G^{(u,u)}$ and $\mathcal D_L^{(u,u)}$, given respectively in
Eqs.~(\ref{eq:DGpole}) and (\ref{eq:DLpole}),
into the matrix equation, Eq.~(\ref{eq:DLvsDG}).
A key observation here is that the contributions proportional to $\ML$ and $\M^s$
do not have poles near the position of the bound state, and thus can be treated
as part of the slowly varying ``background'' underneath the pole.\footnote{%
The factors of $L^3$ convert $\delta_{k' k}$ into $(2\pi)^3 \delta^3(\vec k'-\vec k)$ in
the $L\to\infty$ limit, and do not lead to poles.}
As noted already in the main text, these lead only to higher-order energy shifts.
Thus, when looking for the dominant energy shift we can ignore these terms.
The structure of (\ref{eq:DLvsDG}) then mirrors that of Eq.~(\ref{eq:DGdecomposed}),
and by the same argument as in the first stage we find
\begin{align}
\Delta E(L) &= \Delta E_G(L) + \Delta E_F(L) + \cdots \,,
\\
\Delta E_F(L) &=  \frac{1}{2 E_B}   \frac1{L^3} \sum_{\vec \ell} 
\overline \Gamma^{(u)} \!(\ell) \,  \frac{F^{i \epsilon} (\vec \ell) }{2 \omega_\ell } 
\, \Gamma^{(u)}(\ell)\bigg|_{E=E_B}
   \,.
\label{eq:DEL}
\end{align}
Here $F^{i\epsilon}(\vec \ell)$ is obtained from the matrix version of the same quantity,
Eq.~(\ref{eq:Fdef3}),
by removing the $\delta_{k'k}$:
\begin{equation}
\begin{split}
\label{eq:Fiepsaltdef}
F^{i\epsilon}(\vec \ell)
& \equiv \frac12 \bigg[\frac{1}{L^3} \sum_{\vec k} - \int_{\vec k} \bigg]  \\[-3pt] & \times
\frac{   H(\vec k)
  H(\vec \ell)H(\vec b_{k\ell})}
{2 \omega_k 2 \omega_{k\ell}(E - \omega_k - \omega_\ell -  \omega_{k \ell} + i \epsilon)}
  \,.
  \end{split}
  \end{equation}
We note that the sum over $\vec \ell$ in (\ref{eq:DEL}) arises from
the matrix product in (\ref{eq:DLvsDG}).
We also observe that for the subthreshold energies that we consider here
we can set $\epsilon\to 0$.

Our task is thus to evaluate $\Delta E_F(L)$ and show
that it is suppressed relative to $\Delta E_G(L)$.
We recall from Eq.~(\ref{eq:result}) in the main text that $\Delta E_G(L)$
scales as $e^{-2\kappa L/\sqrt3}/(\kappa L)^{3/2}$.
In the subsequent evaluation we will drop all constants and keep track only of
$L$-dependence.

Substituting the residue factors from Eq.~(\ref{eq:Gammaresult}) we find
\begin{equation}
\Delta E_F(L)   \propto 
\frac{1}{L^3} \sum_{\vec \ell} \left[1 + \frac{3 \ell^2}{4 \kappa^2} \right ]^{-1}   
F^{i\epsilon}(\vec \ell)\bigg|_{E=E_B}
\,.
\end{equation}
The zeta-function can be rewritten using the Poisson summation formula,
following Eqs.~(43) and (C4)-(C6) of Ref.~\cite{thresh}:
\begin{equation}
F^{i\epsilon}(\vec \ell)\bigg|_{E=E_B}
 = -
\frac1{32 \pi m L}  \sum_{\vec s\ne 0} e^{i\pi \vec s\cdot \vec n_\ell} 
\frac{e^{- s L \sqrt{\kappa^2 + 3 \ell^2/4} } }{s}  \,.
\label{eq:FiepsNR}
\end{equation}
Here $\vec s$ is a vector of integers and $\vec n_\ell =L \vec \ell/(2\pi)$.
To obtain this form we have also expanded in powers of $\kappa^2/m^2$ and dropped subleading
contributions. 
The cutoff functions $H$ have also been dropped since they are made redundant by the
natural cutoff in the exponential.

Combining these results, and using the Poisson summation formula on the sum over $\vec \ell$,
we find
\begin{multline}
\Delta E_F(L)  \propto \frac1L  \sum_{\vec n} \int \! \frac{d^3 \ell}{(2 \pi)^3}   
\sum_{\vec s\ne 0} \frac1s \\ 
\times \left[1 + \frac{3 \ell^2}{4 \kappa^2} \right ]^{-1}   
e^{  i L ( \vec n + \vec s/2) \cdot \vec \ell - s L \sqrt{\kappa^2 + 3 \ell^2/4} } \,.
\end{multline}
Evaluating the angular integral leads to
\begin{align}
\begin{split}
\label{eq:InumA}
\Delta E_F(L)  &\propto \frac1{L^2} \sum_{\vec n} \sum_{\vec s\ne 0} \frac1s 
\mathrm{Im}  \int_{- \infty}^{\infty} d\ell  \ell  \\[-10pt] & \hspace{90pt} \times
\frac{e^{L f(\ell)}}{1 + 3 \ell^2/(4 \kappa^2)} \,,
\end{split} \\[5pt]
f(\ell) &= i q - s\sqrt{\kappa^2+3\ell^2/4}\,,
\\
q & = |\vec n + \vec s/2|\,.
\end{align}
The integral can be evaluated by deforming the contour to pass through the
appropriate stationary point, $\ell_0$, and using the steepest descent approximation.
The stationary point and the corresponding exponent are
\begin{align}
\ell_0 &=\frac{2 iq\kappa}{\sqrt{3}} \frac{1}{ \sqrt{3 s^2/4 +  q^2}}\,,
\\
f(\ell_0) &= - \frac{2 \kappa L}{\sqrt3} \sqrt{3 s^2/4 + q^2}\,.
\end{align}
Since the result of the integral
scales as $e^{-L f(\ell_0)}$ it is now clear that the dominant
contributions arise when $s=1$ and $q=1/2$. These come from the 12 terms
having $s = 1$ together with $\vec n=0$ or $\vec n=-\vec s$.\footnote{%
The fact that $\vec n=-\vec s$ contributes equally with $\vec n=0$ implies that
the sum over $\vec\ell$ in the original expression for $\Delta E_F(L)$ cannot
be replaced by an integral, despite the fact that the summand
has no poles in the subthreshold region. If one were to make this replacement
then one would get an answer too small by a factor of two.}
Doing the Gaussian integral we then obtain
\begin{equation}
\Delta E_F(L) \propto  \frac{e^{-2\kappa L/\sqrt3}}{L^{5/2} } + \cdots
\,.
\label{eq:Ifinal}
\end{equation}
As claimed above, this is suppressed by a factor of $1/L$ compared to 
$\Delta E_G(L)$.
 
Note here that we have expanded all quantities, including the poles 
at $\ell= \pm 2 i \kappa/\sqrt{3}$, about the saddle point. 
One might be concerned that this leads to the incorrect scaling since the 
pole at $\ell =  2 i \kappa/\sqrt{3}$ lies close to the stationary point
at $\ell_0 = i \kappa/\sqrt{3}$.
This is not the case, however. 
As one sends $ L \to \infty$ the Gaussian peak becomes
arbitrarily narrow and the effect of the pole is damped away. 
We have checked this result numerically by calculating the ratio of the right-hand sides of
Eqs.~(\ref{eq:InumA}) and (\ref{eq:Ifinal}).
We find that the
quantities indeed asymptote to the same value, although
the convergence is relatively slow, requiring $\kappa L \approx 125$ to
reach subpercent agreement.

\subsection{Dropping $F^{i\epsilon}$ terms for general $\Kdf$}

When $\Kdf\ne 0$,  we have not been able to find a simple expression, akin to
Eq.~(\ref{eq:DLvsDG}), showing the form of contributions proportional to $F^{i\epsilon}$.
Thus we will make the argument that these terms can be dropped in a slightly
different way. This approach is more general and would also work for $\Kdf=0$.
The point is that, in order to obtain the leading-order energy shift from $\MthL^{(u,u)}$,
we can simply drop all terms from this quantity that have subleading volume dependence
when $E=E_B$.
It is not important whether the terms we drop contribute to the energy shift or, say, to a
shift in the residues of the pole. Thus all we need to do is show that terms containing
factors of $F^{i\epsilon}$ are subleading, then it is legitimate to set $F^{i\epsilon}\to 0$
for the purposes of the calculation in the main text.

When we expand out $\MthL$, Eq.~(\ref{eq:M3L}), in powers of $F^{i\epsilon}$,
we find that the latter appears in the forms
\begin{gather}
\begin{split}
\mathcal D^{(u,u)} F^{i\epsilon} \mathcal D^{(u,u)}\,,\ \
\mathcal D^{(u,u)} F^{i\epsilon} \Kdf\,, \\
\Kdf F^{i\epsilon} \mathcal D^{(u,u)} 
\ \ {\rm and}\ \ 
\Kdf F^{i\epsilon} \Kdf
\,.
\end{split}
\label{eq:Fterms}
\end{gather}
For $\Kdf\ne 0$, $\mathcal D^{(u,u)}(\vec k, \vec p)$ will not have the pole form
of Eq.~(\ref{eq:Dpole}). Thus the dependence on its arguments will
not be given by that of $\Gamma( k)\overline \Gamma( p)$.
In particular, there is no reason to expect that the singularity present in $\Gamma$,
Eq.~(\ref{eq:Gammaresult}), will still be present in the dependence of $\mathcal D^{(u,u)}$
on its arguments. Similarly, we do not expect $\Kdf$ to
have any singularities close to threshold, as it corresponds to a quasi-local vertex.
Thus, when all other momenta are held fixed,
we expect the general form of all the terms in Eq.~(\ref{eq:Fterms}) to be
\begin{equation}
\frac1{L^3} \sum_{\vec \ell}g(\vec \ell) F^{i\epsilon}(\vec \ell)\bigg|_{E=E_B}
\,,
\end{equation}
with $g(\vec \ell)$ a smooth function in the threshold region.
Assuming this form, the calculation of the previous subsection shows that
this contribution scales as $L^{-5/2} \exp(-2\kappa L/\sqrt3)$ and is thus subleading.
In fact, the previous calculation shows that this result will hold even if $g(\vec \ell)$ has
a singularity at the same position as that in $\Gamma(\ell)$.

\subsection{Dropping the ``$-1/3$ terms" for general $\Kdf$}

The final task of this appendix is to argue that the $-1/3$ terms in
Eq.~(\ref{eq:LKR}), and implicitly in Eq.~(\ref{eq:KFKfinal}), lead to subleading
volume dependence.
We follow the line of argument used in the previous subsection, namely we work
directly with $\MthL^{(u,u)}$ and do not derive an expression for $\Delta E(L)$.
The volume dependence from the terms of interest arises from the forms
\begin{multline}
\mathcal D^{(u,u)} C^{(-1)} \Kdf \,,\ \
\Kdf C^{(-1)} \mathcal D^{(u,u)}\,,
\\
{\rm and}\ \
\Kdf C^{(-1)} \Kdf
\,.
\label{eq:Cforms}
\end{multline}
These are shorthand for the sum-integral differences, as described in the main text.
For example
\begin{multline}
\mathcal D^{(u,u)} C^{(-1)} \Kdf =
\\
\bigg[\frac1{L^3} \sum_{\vec \ell} - \int_{\vec \ell}\bigg]
\mathcal D^{(u,u)}(\vec p, \vec \ell)
\frac1{2\omega_\ell \M^s(\vec \ell)} \Kdf(\vec \ell, \vec k) \,,
\end{multline}
As noted above, we do not know the momentum dependence
of $\mathcal D^{(u,u)}$ when $\Kdf\ne 0$, and there is no reason to expect
it to have a singularity near threshold. 
For $\Kdf$ we take, as above, a smooth dependence on $\vec \ell$, with no singularities
near threshold. 
Finally, we recall from Eq.~(\ref{eq:1byM2})
that $1/\M^s$ has a branch cut at $\ell^2=-4\kappa^2/3$.
Putting these ingredients together we find that the summand/integrands for all of
the forms in Eq.~(\ref{eq:Cforms}) are expected to have only the
square-root branch cut arising from $1/\M^s$.
This is in contrast to the expression for $\Delta E(L)$, Eq.~(\ref{eq:DE1}),
in which the summand/integrand has an inverse square-root singularity at $\ell^2=-4\kappa^2/3$.

Thus the contributions from the forms (\ref{eq:Cforms}) to $\MthL^{(u,u)}$ are
expected to be proportional to [cf. Eq.~(\ref{eq:introtated})]
\begin{equation}
\frac1L \int_{2\kappa/\sqrt3}^\infty d\ell \ell   e^{-\ell L} \left(\frac{3\ell^2}{4\kappa^2}-1\right)^{1/2}
\propto
\frac{e^{-2\kappa L/\sqrt3}}{L^{5/2}}
\,,
\end{equation}
and are thus suppressed by a power of $1/L$ compared to the leading volume dependence.

\section{Relating residue factors to the Schr\"odinger wavefunction}
\label{app:details}

In this appendix we derive the relation (\ref{eq:Gammafrompsi}) between the
on-shell residue factors $\Gamma^{(u)}$ and $\overline\Gamma^{(u)}$ and the
Schr\"odinger wavefunction $\psi$.

To do so we first relate the Bethe-Salpeter amplitudes of the bound state
to the wavefunction.  Denoting these amplitudes by $\chi$ and $\overline \chi$,
we recall that they are defined via the coefficient
of the bound-state pole in the unamputated $3\to3$ correlation function, 
$S_3$:
\begin{equation}
S_3(p_1',p_2';p_1,p_2;P) \sim \chi(p_1',p_2')  \frac{i}{P^2 - E_B^2}  \overline \chi(p_1,p_2) \,.
\end{equation}
Note that $\chi$ and $\overline\chi$ depend on the 
four-momenta of two of the three particles (the third determined
by energy-momentum conservation). 

The relation between $\chi$ and $\psi$ has been given, 
under certain assumptions, in Ref.~\cite{FF}. In particular the reference assumes that there are
only two-particle, instantaneous interactions and that 
the NR limit has been taken.
Since these are also the assumptions made by MRR,
the results of Ref.~\cite{FF} are sufficient here.
{ The forms of the relations most useful  for our purposes  are}
\begin{multline}
\chi = \beta \sqrt{\tfrac34}\tfrac1m S_1 S_2 S_3 \big[ (E_B-3m-H_0) 
\\
+  \left\{ S_1^{-1} V_{23} S_{23} + S_2^{-1}V_{31} S_{31}
+S_3^{-1} V_{12} S_{12} \right\} \big] \widetilde\psi\,,
\label{eq:chifrompsi3}
\end{multline}
and
\begin{multline}
\overline \chi = 
\beta \sqrt{\tfrac34}\tfrac1m \widetilde \psi^\dagger  \big[ (E_B-3m-H_0) 
\\
+  \left\{ S_{23} V_{23}  S_1^{-1} + S_{31} V_{31}  S_2^{-1}
+ S_{12} V_{12}  S_3^{-1} \right\} \big]  S_1 S_2 S_3 \,,
\label{eq:chibarfrompsi3}
\end{multline}
where we have introduced a normalization factor $\beta$. 
In Appendix~\ref{app:detnorm} we show that $\beta=1$ by 
matching the definitions of $\chi$ and $\psi$ for a 
finite-volume scattering state. 

In Eqs.~(\ref{eq:chifrompsi3}) and (\ref{eq:chibarfrompsi3})
we are using an abbreviated notation that we now explain.
First we note that $\widetilde\psi$ depends on two of the
three momenta, e.g. $\vec p_1$ and $\vec p_2$, with $\vec p_3= -\vec p_1-\vec p_2$.
[or alternatively on the Jacobi momenta as in Eq.~(\ref{eq:psiFT})].
$\chi$ and $\overline\chi$ depend in addition on the energies $E_i$, 
which are constrained to satisfy $E_1+E_2+E_3=E_B$. 
As we show below, only the factor on the first line 
of Eqs.~(\ref{eq:chifrompsi3}) and (\ref{eq:chibarfrompsi3}) 
enters the relation between the wavefunction and the on-shell 
residue factors $\Gamma$, $\overline \Gamma$. 
Note that this factor depends on $H_0$, defined in Eq.~(\ref{eq:H0def}).
$S_i$ are single-particle NR propagators,
\begin{equation}
S_i(E_i,\vec p_i) = \left(E_i-m - \frac{\vec p_i^2}{2m} + i \epsilon\right)^{-1}
\,.
\end{equation}
$S_{ij}$ are two-particle NR propagators that include the potential $V_{ij}$.
In particular, $S_{ij}$ solves the integral equation
\begin{multline}
S_{ij}(E_i\!+\!E_j; \vec p_i, \vec p_j; \vec k_i)
= S^0_{ij}(E_i\!+\!E_j; \vec p_i,\vec p_j) %(2\pi)^3
\\\times %\big[
(2\pi)^3 \delta^3(\vec p_i\!-\!\vec k_i)  
%+ \delta^3(\vec p_j\!-\!\vec k_i)  
%\big]
+    \int \frac{d^3 q_i}{(2\pi)^3} S^0_{ij}(E_i\!+\!E_j;\vec p_i,\vec p_j) 
\\
\times V_{ij}(|\vec q_i-\vec p_i|)
S_{ij}(E_i\!+\!E_j; \vec q_i,\vec q_j; \vec k_i)\,,
\label{eq:Sijdef}
\end{multline}
where $S^0$ is the free two-particle propagator
\begin{align}
\begin{split}
S^0_{ij}(E_i\!+\!E_j;\vec p_i,\vec p_j)^{-1}
&\\[-5pt]
& \hspace{-50pt} = E_i+E_j - 2m - \frac{\vec p_i^2}{2m}- \frac{\vec p_j^2}{2m} + i \epsilon \,,
\end{split}\\
&  \hspace{-50pt} = S_i(E_i, \vec p_i)^{-1}+ S_j(E_j, \vec p_j)^{-1}\,.
\end{align}
We do not show the fourth momentum argument of $S_{ij}$ because
total momentum is conserved, i.e.~$\vec p_i+\vec p_j = \vec q_i+\vec q_j=\vec k_i+\vec k_j$.
Note that Eqs.~(\ref{eq:chifrompsi3}) and (\ref{eq:chibarfrompsi3})
contain implicit three-momentum integrals adjacent to the $S_{ij}$ factors.
Their form can be seen by noting that the shorthand version of Eq.~(\ref{eq:Sijdef})
is
\begin{equation}
S_{ij} = S_{ij}^0 + S_{ij}^0 V_{ij} S_{ij}\,.
\end{equation}
It will be important that the energy dependence of $S_{ij}$ is explicit,
entering only through the $E_i+E_j$ term in $S^0_{ij}$.

As discussed in Ref.~\cite{FF}, with these definitions one can show that $\psi$
satisfies the Schr\"odinger equation if $\chi$ satisfies the Bethe-Salpeter equation
and {\em vice versa}.
We have checked this result.
This does not depend on the overall normalizations 
in Eqs.~(\ref{eq:chifrompsi3}) and (\ref{eq:chibarfrompsi3}),
and in fact we find that a different normalization factor from that given in
Ref.~\cite{FF} is needed in order that  $\psi$ is normalized 
as in Eq.~(\ref{eq:psinorm}).
We explain how we determine the normalization factor, $\beta$, in Appendix~\ref{app:detnorm} below.
First we describe how we proceed from Eq.~(\ref{eq:chifrompsi3})
to the desired result (\ref{eq:Gammafrompsi}).

\subsection{From the Bethe-Salpeter amplitude to $\Gamma^{(u)}$}

To obtain $\Gamma$ from $\chi$ we must amputate and then go on shell,
and in addition multiply by a factor of $-i$ to account for the
overall sign difference in the pole term.\footnote{%
This follows from the fact that amputating $G$ and 
going on shell gives $i \mathcal M_3$, but the relation 
between $\Gamma \overline \Gamma$ and $i \mathcal M_3$ 
differs from that between $\chi \overline \chi$ and $G$ by an overall sign.
The choice of $-i$ rather than $i$ is for convenience.
Note that we must use the same factor 
to relate $\overline\Gamma$ to $\overline\chi$,
i.e. $-i$ and not $i$.}
Amputation requires multiplying by the product of three
relativistic propagators. The relation between the relativistic and the NR propagators near the pole is
\begin{align}
S_{\rm rel}(p_i)^{-1} &= \frac{p_i^2-m^2+ i\epsilon}i 
\\
&\approx \frac{2m}i (E_i-\omega_{p_i} + i\epsilon)
\\
&\approx \frac{2m}i S_i(E_i,\vec p_i)
\,,
\end{align}
with $p_i$ a four-vector.
Thus we find 
\begin{multline}
\Gamma 
= \!\! \lim_{\mathrm{on\ shell}}
(-i) \beta \sqrt{\tfrac34} (8 i m^2) \big[ (E_B- 3m-H_0) 
\\
+ \left\{ S_1^{-1} V_{23} S_{23} + S_2^{-1}V_{31} S_{31}
+S_3^{-1} V_{12} S_{12} \right\} \big] \widetilde \psi\,.
\label{eq:Gammafrompsi3}
\end{multline}

We now argue that the terms involving $S_j^{-1}$ vanish due to the on-shell limit.
We imagine taking this limit by first sending $p_1^2/(2m) \to E_1 - m$, then
$p_2^2/(2m) \to E_2-m$, and finally $p_3^2/(2m) \to E_3-m$. 
The final result must not depend on this choice of ordering.
The first step sets $S_1^{-1}\to 0$, removing the $S_1^{-1}$ term.
The second step similarly removes the $S_2^{-1}$ term.
At this stage we note that $E_B-3m-H_0 = S_3^{-1}$, so it
appears that the two remaining terms on the right-hand side of
Eq.~(\ref{eq:Gammafrompsi3}) are on an equal footing,
and that both vanish when $E_3$ goes on shell.
In fact, the $E_B-3m-H_0$ term does not vanish, as we show in the main text by
explicit calculation. This is due to a corresponding divergence in $\widetilde\psi$.
This divergence does not save the $S_3^{-1}$ term from vanishing, however,
because of the momentum integral that implicitly accompanies the factor of $S_{12}$. 
This integral remains even when the external momenta are set on shell,
and does not diverge.
Thus we find
\begin{align}
\Gamma 
= \!\! \lim_{\mathrm{on\ shell}}
\beta 4\sqrt3 m^2 (E_B- 3 m -H_0) \widetilde\psi \,, %\Big|_{\rm on\ shell}\,,
\label{eq:Gammafinal3}
\end{align}
a result that indeed is independent of the manner in which we approach the on-shell point.
Similarly we find
\begin{align}
\overline \Gamma 
= \!\! \lim_{\mathrm{on\ shell}}
\beta 4\sqrt3 m^2 \widetilde\psi^\dagger(E_B-3m-H_0) \,. % \Big|_{\rm on\ shell}\,.
\end{align}
We note that, up to overall normalization factors, the same expression holds
for the relation of $\Gamma$ to $\psi$ in the two-particle case.

The final step is to argue that we obtain the unsymmetrized residue $\Gamma^{(u)}$
by replacing $\widetilde\psi$ with $\widetilde\phi_3$ in Eq.~(\ref{eq:Gammafinal3})
[and similarly for $\overline\Gamma^{(u)}$].
This leads to the desired result (\ref{eq:Gammafrompsi}).
First, we note that this claim is consistent with Eq.~(\ref{eq:Gammafinal3}).
This is because the full wavefunction is given 
by summing the three components related by permutations
\begin{equation}
\widetilde \psi = 
\widetilde\phi_1(\vec k_{23}, \vec k_1) +
\widetilde\phi_2(\vec k_{31}, \vec k_2) +
\widetilde\phi_3(\vec k_{12}, \vec k_3) 
\,,
\end{equation}
while the full $\Gamma$ is obtained by similarly symmetrizing $\Gamma^{(u)}$.
Second, we use the observation given in the main text,
namely that, if we imagine iteratively solving the Faddeev equation (\ref{eq:Fadeev3}) and
its permutations, we obtain for $\widetilde \phi_3$
a sequence of contributions in which the first interaction is always 
between particles 1 and 2. This is precisely the definition of the
unsymmetrized amplitude $\Mth^{(u,u)}$, from which $\Gamma^{(u)}$ is obtained.

\subsection{Deriving the normalization factor}

\label{app:detnorm}

We have found that the simplest way to determine the overall normalization of Eqs.~(\ref{eq:chifrompsi3}) and (\ref{eq:chibarfrompsi3}), i.e.~the value of $\beta$,
is to use a somewhat indirect method.\footnote{%
In principle, one should be able to use the normalization equation satisfied by the
Bethe-Salpeter amplitude as well as that satisfied by the Schr\"odinger wavefunction
to deduce the desired normalization factor, 
but we have not been able to complete the calculation in this manner
due to the complicated form of Eq.~(\ref{eq:chifrompsi3}).}
We consider the poles in the finite-volume $3\to3$ correlation function,
for which we can directly calculate both $\chi$ and $\psi$.
The derivation of Eq.~(\ref{eq:chifrompsi3}) relies on $\chi$ satisfying the
Bethe-Salpeter equation and $\psi$ the Schr\"odinger equation, both of which
remain valid in finite volume. The only change is that momentum integrals become sums, 
but if we work in large volumes this difference is a subleading effect. 
The motivation of studying a finite-volume correlator is that this has 
an infinite tower of poles, and any one of these can be used to study 
the relation between $\chi$ and $\psi$.
The derivation of this relation
does not rely on the pole in the correlator corresponding to a bound state.
It can equally well be a finite-volume scattering state, 
as long as it is near enough to threshold to be in the nonrelativistic regime. 

Thus our idea is to use the results of Ref.~\cite{HSpert}, in which we did a perturbative
calculation of the $3\to3$ correlation function in finite-volume in $\lambda\phi^4$ theory.
Since the relations we are testing are essentially kinematical, we can work here at
infinitesimal $\lambda$, and keep only the lowest term in the expansions of the relevant
quantities. 
The relevant correlator is\footnote{%
We have checked our method by repeating the calculation for two particles, and finding
the correct relation between $\chi$ and $\psi$ in that case.}
\begin{equation}
C_{3}(\tau) = \langle \widetilde\phi_{\vec 0}(\tau)^3 \widetilde\phi_{\vec 0}(0)^3\rangle
\,,
\end{equation}
where $\widetilde\phi_{\vec 0}(\tau)$ is the zero-spatial-momentum field at Euclidean time $\tau$.
We focus on the contribution of the state nearest threshold,
\begin{equation}
C_{3}(\tau) \supset Z_{3} e^{-( 3m + \Delta E_3) |\tau|}\,.
\end{equation}
What we need from Ref.~\cite{HSpert} are the results
\begin{equation}
Z_3 = \frac{3! L^{9}}{(2m)^3} \left[1 + {\cal O}(\lambda/L^3) \right]
\quad{\rm and}\quad
\Delta E_3 = {\cal O}(\lambda/L^3)
\,.
\label{eq:Znres}
\end{equation}

We also need the form of the wavefunction for this state, 
or more precisely (as we will see) the
momentum-space wavefunction at vanishing momenta.
At leading order the state simply consists of three free particles
in a cubic box of size $L^3$ each with zero momentum.
It follows that the position-space wavefunction is a constant, 
$\psi(\vec x_3,\vec y_3) = c_3$
and this constant can be determined from the normalization condition, Eq.~(\ref{eq:psinorm})
(with here $A=1$).
One can rearrange the fundamental domain for three particles such that the period in
each component of the Jacobi coordinates $\vec x_3$ and $\vec y_3$ is
$L$ and $2 L/\sqrt3$, respectively.
Using this we have
\begin{equation}
\begin{split}
\int d^3x_3 d^3y_3 J |\psi|^2 = 6 \ \ & \Longrightarrow\ \
\frac{J}{J} |c_3|^2 L^6 = 6 \,, \\
\ \ & \Longrightarrow\ \  |c_3| = \frac{\sqrt6}{L^3} \,.
\end{split}
\end{equation}
Thus we find
\begin{equation}
\widetilde\psi_3(\vec 0,\vec 0)
= \int d^3 x_3 d^3 y_3 J \psi(\vec x_3,\vec y_3)
= c_3 L^6 \,.
\label{eq:psi3zeromom}
\end{equation}

Our next step is to Fourier transform $C_3(\tau)$ in time, 
so that it becomes the momentum-space correlator used to define $\chi$ and $\overline\chi$:
\begin{align}
\int dt\; e^{i P^0 t} C_3(t) 
&=
\int (-i) d\tau e^{ P^0\tau} C_3(\tau) \,,
\\
&\sim \frac{-i Z_3 2 (3m+\Delta E_3)}{(3m + \Delta E_3)^2 - (P^0)^2} %+ \textrm{non-pole}
\,,
\label{eq:intCA}
\end{align}
where the $\sim$ indicates that the two sides differ by terms that are finite at the pole.
In the first step we have analytically continued to Euclidean time;
in the second, we evaluate the integral assuming $P^0 < (3m + \Delta E_3)$
and then analytically continue to general $P^0$.

Alternatively, one can evaluate the integral in terms of the off-shell
momentum-space $3\to3$ correlator $S_3(k'_{12},k'_3; k_{12},k_3; P)$, 
where we 
have used the Jacobi momenta (\ref{eq:Jacobimomenta}) extended to four-vectors,
and $P^\mu=(P^0,\vec P)$ is the total four-momentum.
Standard manipulations lead to
\begin{multline}
\int dt\; e^{i P^0 t} C_3(t) 
 =
L^3 \int \frac{dk_{12}^0}{2\pi} \int \frac{dk_{3}^0}{2\pi} 
\int \frac{dk'^{0}_{12}}{2\pi} \int \frac{dk'^{0}_{3}}{2\pi} 
\\
\times
S_{3}(k'_{12},k'_3;k_{12},k_3;P)
\bigg|_{\vec k_{12}=\vec k_3=\vec k'_{12}=\vec k'_3=\vec P=\vec 0} \,.
\end{multline}
Using the definition of the Bethe-Salpeter amplitudes at the pole
\begin{multline}
S_3(k'_{12},k'_3; k_{12},k_3; P) \sim %= 
\\
\chi(k'_{12},k'_3) \frac{i}{P^2 - E_{\rm pole}^2}
\overline\chi(k_{12},k_3) \,, % + \textrm{non-pole}\,,
\label{eq:BSampdef}
\end{multline}
we find
\begin{align}
\int dt e^{i P^0 t} C_3(t) 
&\sim
L^3  \frac{iX  \overline X}{(P^0)^2 - E_{\rm pole}^2}  \,, % + \textrm{non-pole}
\label{eq:intC3B}
\end{align}
where
\begin{align}
X &=  \int \frac{dk_{12}^0}{2\pi}  \frac{dk_3^0}{2\pi} \chi([k_{12}^0,\vec 0],[k_{3}^0,\vec 0])  \,,
\label{eq:Xdef}
\\
\overline X
&=  \int \frac{d{k}^0_{12}}{2\pi}  \frac{d{k}^0_3}{2\pi} 
\overline \chi([{k}_{12}^0,\vec 0],[{k}^0_{3},\vec 0])  \,.
\label{eq:Xbardef}
\end{align}
Comparing to Eq.~(\ref{eq:intCA}), and keeping
the leading terms in perturbation theory for $Z_3$ and $\Delta E_3$, we find
\begin{equation}
X \overline X = \frac{6 Z_3m}{L^3} = \frac{36 L^6}{8m^2}
\,.
\label{eq:XbarXres}
\end{equation}

We are finally ready to determine the normalization factor $\beta$ in
Eqs.~(\ref{eq:chifrompsi3}) and (\ref{eq:chibarfrompsi3}).
Replacing $E_B\to E_{\rm pole} = 3 m + \Delta E_3$, 
we substitute the wavefunction (\ref{eq:psi3zeromom}) 
to deduce the values of $\chi$ and $\overline \chi$ 
predicted by these relations. 
To simplify the result, note that we can evaluate the 
single-particle propagators at $E_{\rm pole}$ as well 
as vanishing spatial momenta
\begin{align}
S_1^{-1} &= \frac{\Delta E_3}{3}+ k_{12}^0 + \frac{k_3^0}{2} + i\epsilon\,,
\\
S_2^{-1} &= \frac{\Delta E_3}{3}-k_{12}^0 + \frac{k_3^0}{2} + i\epsilon\,,
\\
S_3^{-1} &= \frac{\Delta E_3}{3} - k_3^0 + i\epsilon\,.
\end{align}
We now evaluate the integrals,
implicit in (\ref{eq:chifrompsi3}) and (\ref{eq:chibarfrompsi3}), 
and find that it is always possible to close the
contour such that only the 
$E_{\mathrm{pole}}-3m-H_0$ term contributes. 
For example, for the $S_3^{-1}$ term one can close the $k_{12}^0$ contour below
and pick up the pole in $S_1$, 
but the remainder can be written as a some of the terms
containing powers of $S_{12}^0$ 
[as can be seen by iterating Eq.~(\ref{eq:Sijdef})].
All these terms have the $k_3^0$ pole below the axis, and so vanish when
we close the $k_3^0$ contour above.
Evaluating the integrals for the $E_{\mathrm{pole}}-3m-H_0$ term, we find
\begin{equation}
X = (-i)^2 \beta \sqrt{\tfrac34}\tfrac1m \widetilde \psi(\vec 0, \vec 0)
\,.
\end{equation}
The same expression holds for $\overline X$ in terms of $\widetilde\psi^\dagger$.

Thus we deduce that the value of $X \overline X$ {determined from}
Eqs.~(\ref{eq:chifrompsi3}) and (\ref{eq:chibarfrompsi3}) is 
\begin{equation}
X \overline X = \beta^2 \frac34 \frac1{m^2} |\widetilde \psi(\vec 0, \vec 0)|^2 = \beta^2
\frac{3|c_3|^2 L^{12}}{4m^2} = \beta^2 \frac{9L^6}{2m^2} \,,
\end{equation}
where we have used that $\beta$ is assumed to be a positive, real number. Comparing this to the direct evaluation, Eq.~(\ref{eq:XbarXres}), we deduce $\beta=1$ as claimed.

\section{An identity for the Schr\"odinger wavefunction}
\label{app:identity}

In this appendix we use the explicit form of $\phi$,
Eq.~(\ref{eq:phidef}), to derive the identity Eq.~(\ref{eq:identity}).
We first reproduce the identity 
\begin{align}
\left( -{\kappa^2}  + \vec \nabla_{x_3}^2  + \vec \nabla_{y_3}^2\right) \phi(R, \alpha_3) 
&= m h(|\vec y_3|) \delta^3(\vec x_3) \,,
\label{eq:identity2}
\\
h(y) &= -4 \pi \frac{b}{m}   \frac{K_{i s_0}(\kappa y)}{y}\,,
\label{eq:hres}
\\
b &= A \kappa \sqrt{D_0} \,,
\end{align}
and the form of $\phi$
\begin{align}
\phi(R, \alpha) &= b \frac{K_{is_0}(\sqrt2 \kappa R)}{R^2}
\frac{\sh(s_0[\pi/2-\alpha])}{\sh(s_0 \pi/2)} \frac1{\sin (2\alpha)}
\,,
\end{align}
where, as above, $\sh \, x = \sinh x$. 

Writing the Laplacian in hyperspherical coordinates (as described, for example, in Ref.~\cite{BH})
one easily verifies that the left-hand side of Eq.~(\ref{eq:identity2}) vanishes except 
at the end points $R=0$ and $\alpha_3=0$, where $\phi$ diverges.
To study these singular points it is better to use the coordinates $\vec x_3$ and $\vec y_3$.
Given the definition of $R$, Eq.~(\ref{eq:Rhyper}),
$R$ vanishes only when both $\vec x_3$ and $\vec y_3$ vanish, 
i.e. when all three particles are at the same position.\footnote{%
$K_{is_0}(z)$ itself has an indeterminate limit at $z=0$ since the function 
remains finite but oscillates as a function of $\ln z$---see Eq.~(\ref{eq:Kapprox}).
The  divergence at $R=0$ occurs because of the $1/R^2$ factor in $\phi$.}
By contrast, $\alpha_3$, defined in Eq.~(\ref{eq:alphahyper}),
vanishes when $\vec x_3=0$ for any finite $|\vec y_3|$, 
i.e when particles 1 and 2 are coincident.
We conclude that the left-hand side of
Eq.~(\ref{eq:identity2}) vanishes except when $\vec x_3=0$,
and thus that the identity holds for $\vec x_3 \ne 0$.
We also note that, since $\phi$ depends only the magnitudes of $\vec x_3$ and
$\vec y_3$, and given that this property is maintained by the operator on the
left-hand side of Eq.~(\ref{eq:identity2}),
the function $h$ can only depend on the magnitude of $\vec y_3$, as shown.

To check the ansatz (\ref{eq:identity2}) also at $\vec x_3=0$, we proceed in two stages.
First, we fix $\vec y_3$ to a nonzero value, and send $r=|\vec x_3|/|\vec y_3| \to 0$.
Expanding $\phi$ in this regime, and using $|\vec x_3| = \sqrt2 R \sin\alpha_3$, we find
\begin{equation}
\phi(R,\alpha_3) = \frac{m}{4\pi} \frac1{|\vec x_3|}  h(|\vec y_3|) 
\left[1 + \mathcal O(r)\right]
\,.
\label{eq:phirzero}
\end{equation}
The operator on the left-hand side of Eq.~(\ref{eq:identity2}) gives a finite result
(in fact, zero) when acting on this form except for 
\begin{equation}
\vec\nabla_{x_3}^2 \frac1{|\vec x_3|} = -4\pi \delta^3(\vec x_3)
\,.
\end{equation}
Thus one finds the right-hand side of the identity.\footnote{%
To check this one can integrate both sides of the equation
over a three-dimensional ball in $\vec x_3$ of radius $\epsilon$ with $\epsilon\to 0^+$.
The integrals of the two sides indeed agree.}

The second stage is to consider the region where $\vec x_3\to 0$ with $r$ fixed,
so that both $\vec x_3$ and $\vec y_3$ are vanishing.
Then the approximation of Eq.~(\ref{eq:phirzero}) does not apply,
and the issue is whether there could be an additional term on the right-hand side
of Eq.~(\ref{eq:identity2}) proportional to the six-dimensional
delta-function $\delta^3(\vec x_3)\delta^3(\vec y_3)$.
Such a term would not have contributed in the first stage of the argument.
To address this possibility we integrate both sides of (\ref{eq:identity2}) over a six-dimensional
ball of radius $\sqrt{\vec x_3^2+\vec y_3^2}=\epsilon$, with $\epsilon \kappa \ll 1$. 
A $\delta^3(\vec x_3)\delta^3(\vec y_3)$ term would then lead to an additional constant,
so that the results from integrating the two sides of (\ref{eq:identity2}) would not agree.
In fact, we find that the results do agree, as we now show.

The integral over the right-hand side gives
\begin{equation}
I_R = -(4\pi)^2 b \int_0^{\epsilon} y K_{is_0}(\kappa y) dy\,.
\label{eq:IRHS}
\end{equation}
To evaluate this we use the small argument form of the Bessel function
\begin{equation}
K_{is_0}(z) \approx  a_1 \sin(s_0 \ln z\! +\! a_2)  \qquad (0 < z \ll 1)\,,
\label{eq:Kapprox}
\end{equation}
where $a_1$ and $a_2$ are real constants whose values we will not need.
Then one finds
\begin{multline}
I_R = - \epsilon^2 (4\pi)^2 b a_1 \frac1{4+s_0^2} 
\bigg\{2 \sin[s_0\ln(\epsilon\kappa)\!+\!a_2]
\\ 
- s_0 \cos[s_0\ln(\epsilon\kappa)\!+\!a_2]\bigg\} + \cdots\,,
\label{eq:IRres}
\end{multline}
where the ellipsis indicates terms of higher order in $\epsilon$.

The integral over the left-hand side of (\ref{eq:identity2}) breaks into two parts.
The first comes from the $\kappa^2$ term and is easily found to scale as $\epsilon^4$,
and thus can be dropped.
The second comes from the action of the 
six-dimensional Laplacian, and can be rewritten using the
six-dimensional divergence theorem as
\begin{align}
I_L &= \oint \hat e_R \cdot \vec \nabla \phi 
= \oint \frac1{\sqrt2} \frac{\partial \phi}{\partial R}\Bigg|_{R=\epsilon/\sqrt2}
\,.
\end{align}
Here the integral is over the surface of the ball, 
$\hat e_R$ is the hyperradial unit vector,
and to obtain the second form we have used $\vec x_3^2+\vec y_3^2 = 2 R^2$.
Using the integration measure in hyperspherical coordinates~\cite{BH}, the integral becomes
\begin{equation}
I_L =  (4\pi)^2 \sqrt2 \left(\frac{\epsilon}{\sqrt2}\right)^5
\int d\alpha \sin^2(2\alpha) 
\frac1{\sqrt2} \frac{\partial \phi}{\partial R}\Bigg|_{R=\epsilon/\sqrt2}
\,.
\end{equation}
This evaluates to the same result (\ref{eq:IRres}) as $I_R$, 
thus completing this check.

{
Another possibility for additional terms 
on the right-hand side of Eq.~(\ref{eq:identity2})
is that there could be derivatives of a six-dimensional  delta-function.
There is some reason to expect this for radial derivatives because
$K_{is_0}(z)$ oscillates increasingly rapidly as $z\to 0$,
as shown by Eq.~(\ref{eq:Kapprox}). The dependence on $\alpha$, however,
is much smoother, so we do not expect derivatives with respect to $\alpha$
to occur.
Terms with radial derivatives acting on a delta-function 
can be ruled out as follows: 
integrate the two sides of Eq.~(\ref{eq:identity2})
over the same ball as used above,
but now using the weight functions $R^n$ (with $n>0$). If the two sides match, then such derivative terms must be absent. 
We have verified that indeed, for this class of weight functions,
the integrals of the two sides of Eq.~(\ref{eq:identity2}) agree.}

\bibliographystyle{apsrev4-1} %%% physical review
\bibliography{ref} %%% ref.bib file

\end{document}